\documentclass[a4paper,11pt]{article}
\pdfoutput=1 

\usepackage{jinstpub} 
 
\usepackage{hyperref}
\usepackage{xspace}
\usepackage{color}

\title{\boldmath Evaluation of the Constant Fraction Time-Over-Threshold (CF-TOT) method for neutron-gamma pulse shape discrimination}

\author[a,b,1]{A. Roy,\note{Corresponding author.}}
\author[b]{D. Vartsky,}
\author[c]{I. Mor,}
\author[b,d]{E. O. Cohen,}
\author[b,d]{Y. Yehuda-Zada,} 
\author[d]{A. Beck} 
\author[a]{and L. Arazi.}

\affiliation[a]{Unit of Nuclear Engineering, Ben-Gurion University of the Negev, Beer-Sheva, Israel}
\affiliation[b]{Department of Particle Physics and Astrophysics, Weizmann Institute of Science, Rehovot, Israel}
\affiliation[c]{Soreq Nuclear Research Center, Yavne, Israel}
\affiliation[d]{Nuclear Research Centre Negev, Beer-Sheva, Israel}

\emailAdd{arindam@post.bgu.ac.il}

\abstract{The use of Time-over-Threshold (TOT) for the discrimination between fast neutrons and gamma-rays is advantageous when large number of detection channels are required due to the simplicity of its implementation. However, the results obtained using the standard, Constant Threshold TOT (CT-TOT) are usually inferior to those obtained using other pulse shape discrimination (PSD) methods, such as Charge Comparison or Zero-Crossing approaches, especially for low amplitude neutron/gamma-ray pulses. We evaluate another TOT approach for fast neutron/gamma-ray PSD using Constant-Fraction Time-over-Threshold (CF-TOT) pulse shape analysis. The CT-TOT and CF-TOT methods were compared quantitatively using digitized waveforms from a liquid scintillator coupled to a photomultiplier tube as well as from a stilbene scintillator coupled to a photomultiplier tube and a silicon photomultiplier. The quality of CF-TOT neutron/gamma-ray discrimination was evaluated using Receiver Operator Characteristics curves and the results obtained with this approach were compared to the that of the standard CT-TOT method. The CF-TOT PSD method results in $>99.9\%$ rejection of gamma-rays with $>80\%$ neutron acceptance, much better than CT-TOT.}

\begin{document}
\maketitle
\flushbottom

\section{Introduction}
\label{sec:intro}

The standard and most widely accepted pulse shape discrimination methods used for particle identification are Charge Comparison (CC)~\cite{knoll2010radiation,Moszynski:1994ag} and Zero-Crossing (ZC)~\cite{4324923,SPERR197455}. Both methods have been implemented by using suitable electronic circuits or by directly digitizing the detector current pulse and performing subsequent analysis~\cite{SOSA201672,NAKHOSTIN2010498}.
In systems where hundreds to thousands of detection channels must be used, such as for positron emission tomography (PET)~\cite{Roncali2011,OTTE2006417}, high energy physics detectors~\cite{Garutti_2011,SIMON201985}, multiple-channel fast neutron tomography detectors~\cite{ADAMS20161}, or systems for explosives detection by Gamma-Ray Resonant Absorption (GRA)~\cite{10.1117/12.508262}, there is a need for a simple and rapid analysis of pulse-height and pulse shape, without digitization of the full waveform. In the past two decades there was a significant effort in developing fast methods for pulse-height analysis (PHA) and pulse shape discrimination (PSD). One of the fastest and simplest to implement is the time-over-threshold (TOT) method, credited to D. Nygren~\cite{603658,Nygren91}.\\

\begin{figure}[!ht]
\centering
\includegraphics[width=.8\textwidth]{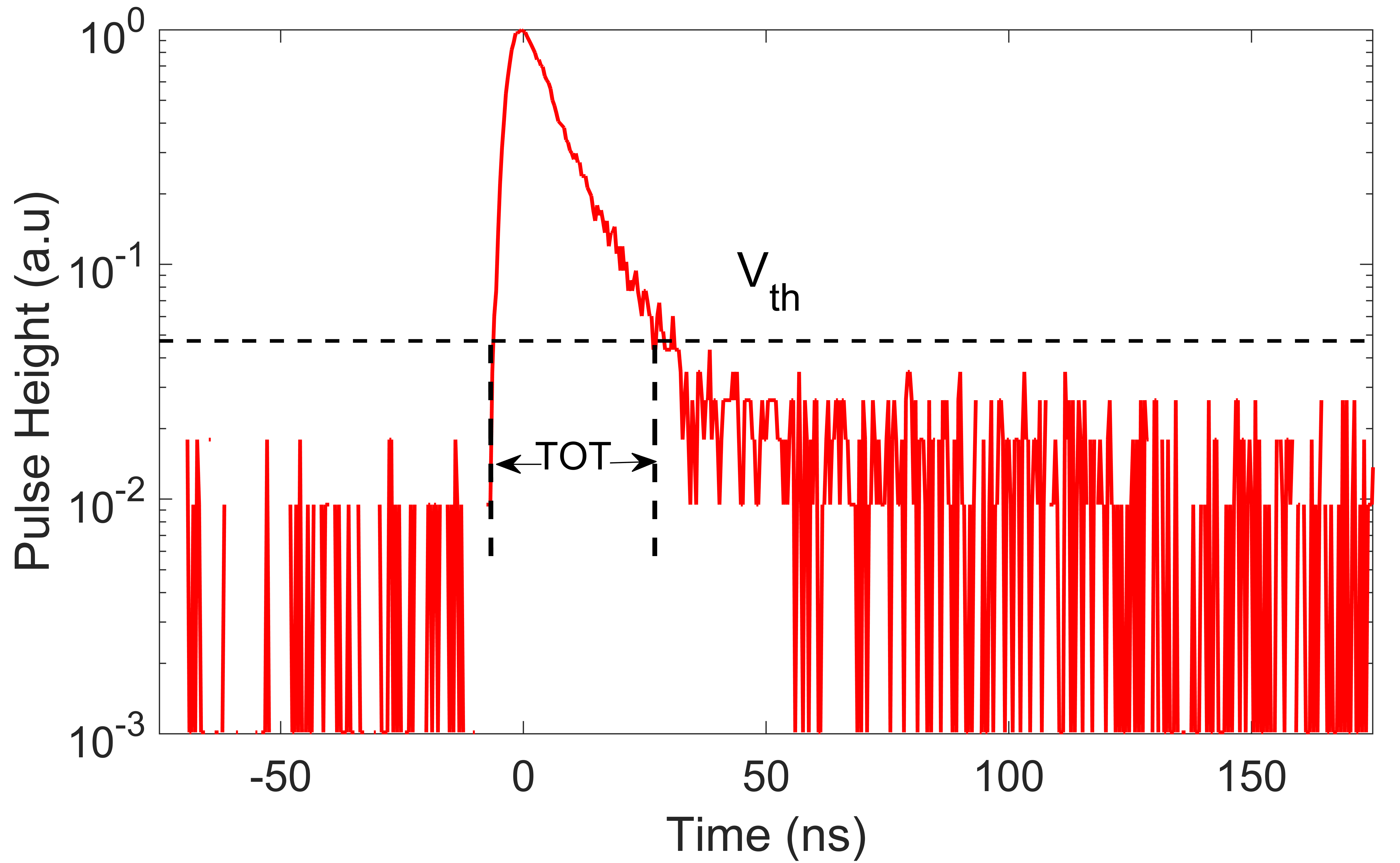}
\caption{\label{fig:i} Illustration of the time over threshold (TOT) extraction for a PMT reading the liquid scintillation signal of a $^{137}$Cs induced event.}
\end{figure}

\noindent TOT is defined as the time interval during which a detected pulse exceeds a constant threshold. Figure \ref{fig:i} illustrates the concept of TOT for an anode pulse obtained from an organic liquid scintillator (NE213 type) coupled to a photomultiplier tube (PMT), exposed to gamma-rays from a $^{137}$Cs source.
For a constant threshold V$_{th}$, and for a pulse of a constant shape, the TOT increases with the pulse amplitude, in a logarithmic manner. For a varying pulse shape, TOT is dependent on both the pulse amplitude and the shape. The TOT method has several advantages for pulse-height analysis:

\begin{itemize}
\item Simplicity of implementation
\item Low power consumption
\item Well suited for multi-channel readout systems in pixelated detectors using Application Specific Integrated Circuit (ASIC). Indeed, several companies manufacture multi-channel ASIC electronics for Time-of-Flight-PET~\cite{ORITA2018303,DULSKI2021165452}, which incorporate TOT features.
\item Less prone to pile-up.
\end{itemize}

\noindent Some of the disadvantages of the TOT method are:
\begin{itemize}
\item Non-linear relationship between TOT and the input pulse charge Q
\item Limited dynamic range
\item Results influenced by noise and local pulse irregularities, such as reflections and after-pulses.
\end{itemize}

\noindent As the relationship between TOT and pulse amplitude or charge is strongly non-linear, several authors proposed various methods for obtaining a linear TOT-charge relationship, such as dynamic time-over-threshold (DToT)~\cite{6154440,ORITA2015154}, time-over-linear-threshold (TOLT)~\cite{https://doi.org/10.48550/arxiv.1806.02494}, and multiple-thresholds (MToT)~\cite{Amiri2015,https://doi.org/10.48550/arxiv.1702.01066}.\\

\noindent The TOT method has been investigated also for applications such as Time-of-Flight-PET~\cite{ORITA2018303,DULSKI2021165452}, PSD in phoswich detectors~\cite{Chang_2016},  and PSD in $^{4}$He gas scintillation counters~\cite{2018EPJWC.17007002B}. Ota proposed a dual time-over-threshold method for the estimation of both energy and scintillation decay time in LYSO detectors~\cite{Ota}, assuming a single pulse decay time constant. In addition, considerable work has been performed on neutron/gamma-ray discrimination in organic liquid and solid scintillators~\cite{JASTANIAH2004202,1351874}. With respect to neutron and gamma-ray PSD by TOT, Jastaniah et al~\cite{JASTANIAH2004202} observed that for low energy neutron and gamma-ray events, the TOT method was not satisfactory due to the overlap between the neutron and gamma-ray TOT. Amiri et al~\cite{Amiri2015} proposed a method termed “Distance Based Method”, in which the digitized pulses were normalized to their peak amplitudes and the TOT was determined at a selected percentage of the pulse amplitude, thus improving the TOT PSD performance.\\

\noindent The aim of this paper is to perform a comprehensive quantitative comparison between the conventional TOT PSD method which uses a fixed threshold on all pulses (CT-TOT) and the constant fraction threshold TOT (CF-TOT) method by an off-line analysis of recorded fast-neutron and gamma-ray waveforms. Our ultimate goal is to avoid pulse digitization and perform multichannel CF-TOT PSD without determining pulse peak amplitude, using a simple analogue circuitry, suitable for multichannel ASIC implementation.

\section{TOT pulse shape discrimination}

\begin{figure}[!ht]
\centering
\includegraphics[width=.86\textwidth]{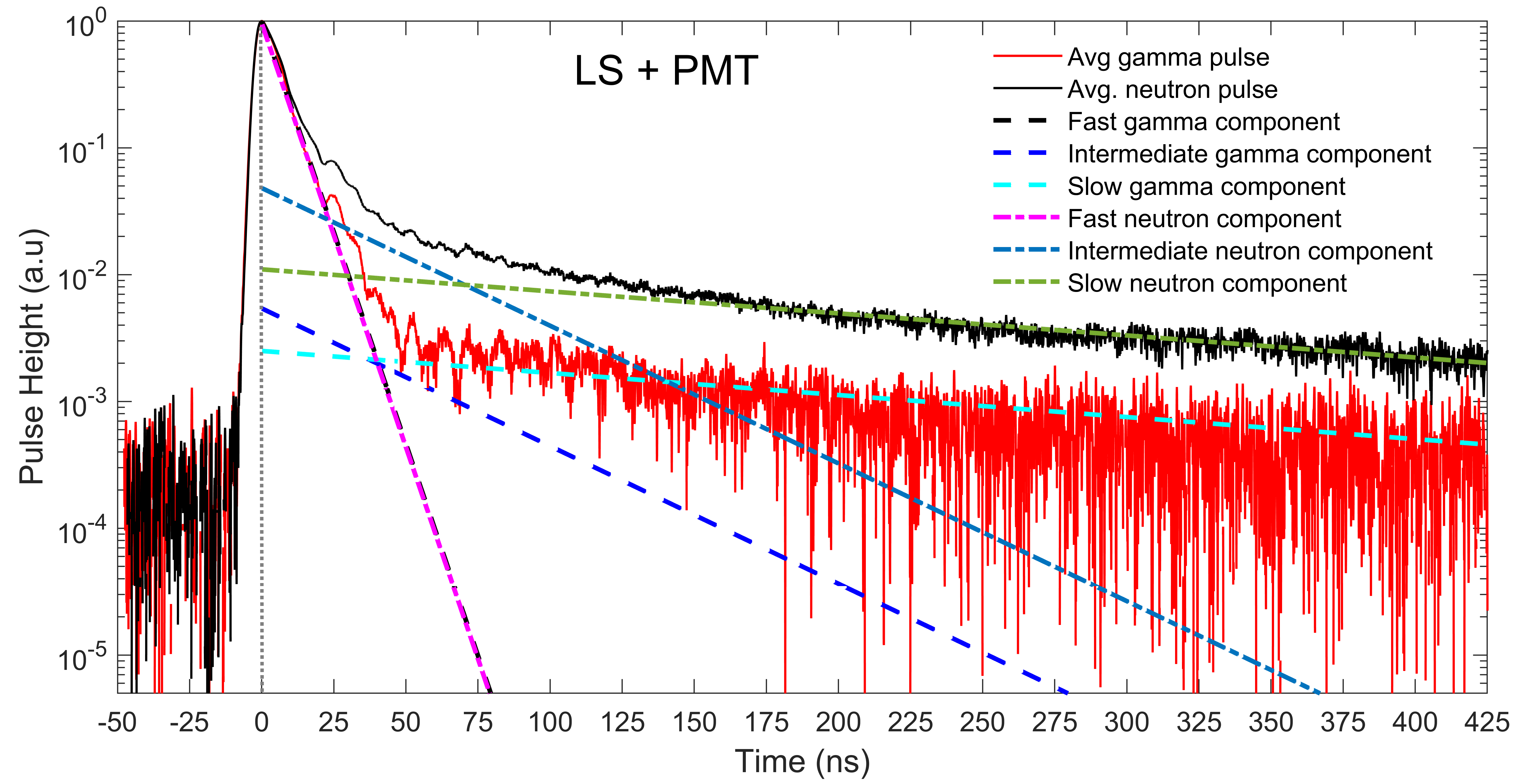}
\caption{\label{fig:ii} Average neutron and gamma-ray pulses, normalized to peak amplitude, obtained using a liquid scintillator and a PMT. The straight lines indicate the contribution of each component to the pulse. The fast component of the neutron and the gamma ray pulse are basically identical and hence the straight line indicating the fast components of the two pulses overlap.}
\end{figure}

\noindent Figure \ref{fig:ii} shows the normalized average neutron and gamma-ray current pulses obtained from a NE-213/BC501 type liquid scintillator coupled to a PMT, by irradiating the detector with an AmBe source. The pulses are shown on a logarithmic scale to illustrate the differences in the tail shape. The current pulse is characterised by a fast rise time and a slow decay. The trailing side of the pulse can be decomposed into three exponentials with a fast, intermediate, and slow time constant~\cite{5594662}, as illustrated in Figure \ref{fig:ii}. These time constants are similar for proton recoils (induced by fast neutrons) and electrons (induced by gamma-rays), but the proportions of the amplitude for each pulse component are different for the two particles. The same applies to some plastic scintillators, organic glass scintillators and stilbene crystals~\cite{Kobylka}. For our liquid scintillator-PMT configuration, the calculated decay times are 6 ns, 37 ns and 250 ns for the fast, intermediate and the slow components, respectively. The magnitude of the fast component is very similar for both types of particles, but the contribution of the slow component, which is responsible for the difference between neutrons and gamma-rays, is $>$3-4 times larger for neutrons.\\

\begin{figure}[!ht]
\centering
\includegraphics[width=.8\textwidth]{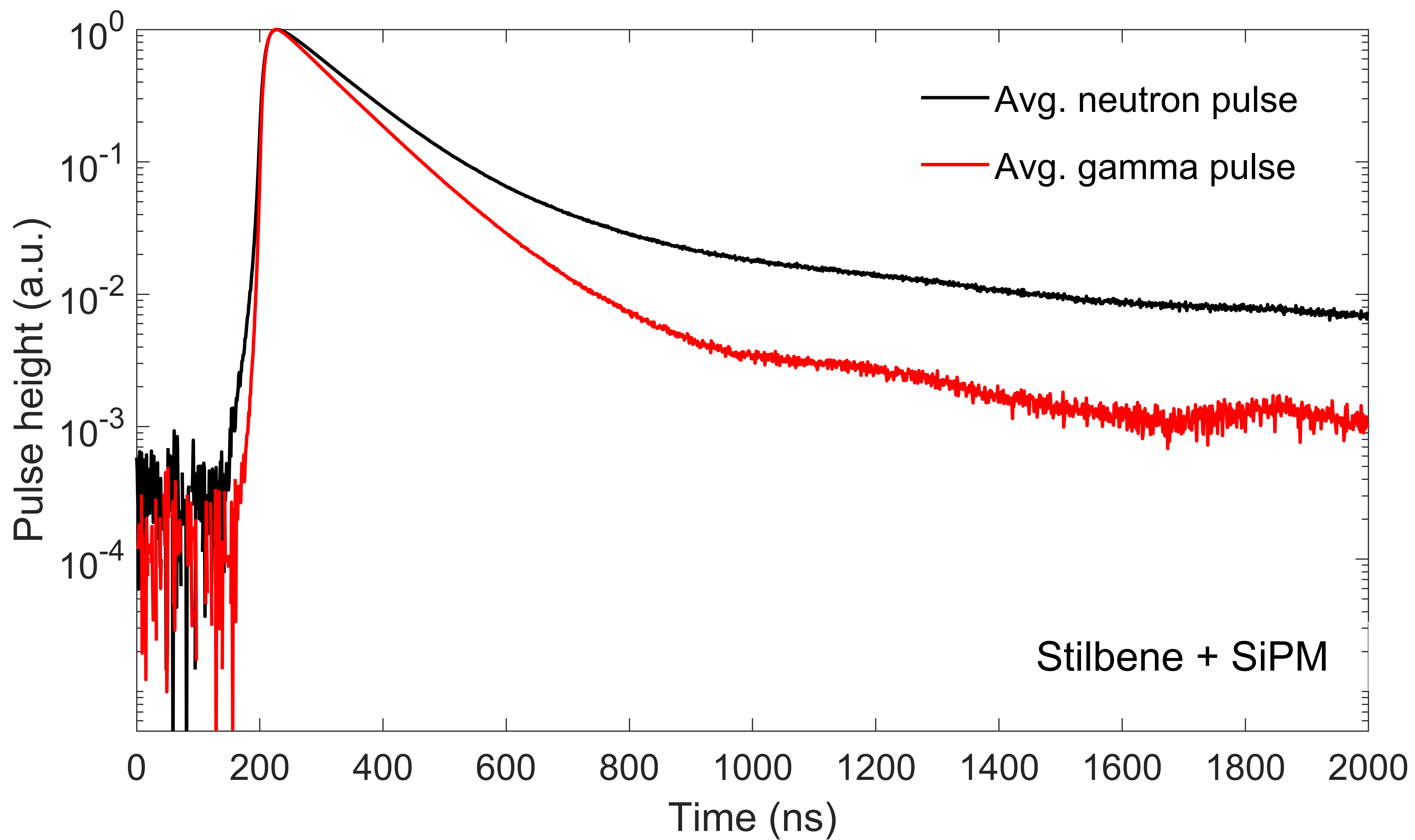}
\caption{\label{fig:iii} Average neutron and gamma-ray pulses, normalized to peak amplitude, obtained using a stilbene coupled to a quad SiPM (Hamamatsu Quad VUV4 MPPC, model S13371-6050CQ-02, 2x2 array of size 6$\times$6 mm$^{2}$, spaced 0.5 mm apart)~\cite{Hamamatsu1,Hamamatsu2}. }
\end{figure}

\noindent Figure \ref{fig:iii} shows the average neutron and gamma-ray pulses, normalized to peak amplitude, obtained for a stilbene crystal coupled to a silicon photomultiplier (SiPM). As the single-electron response of our SiPM is much slower than that of the PMT ($\sim$15 ns rise time and 150 ns fall time), the resulting pulse is much slower and smoother than that of the PMT.

\subsection{Constant threshold TOT (CT-TOT)}

The standard TOT analysis is performed using a constant threshold. In order to discriminate between neutrons and gamma-rays we must sample mainly the slow pulse component, thus the threshold level should be chosen as low as possible, just above noise. This is not always possible due to the varying features of the pulse as can be seen in Figure \ref{fig:ii}. Depending on the pulse magnitude relative to the constant threshold, a different component is sampled for each pulse. For relatively low-amplitude pulses, one samples mainly the fast component, whose magnitude is not significantly different for neutrons and gamma-rays. Thus, for low-energy pulses, one does not expect satisfactory separation between the different particles. As the noise varies for each type of scintillator/light sensor combination, the selected pulse-height threshold represents an optimal choice for a given configuration.

\subsection{Constant fraction TOT (CF-TOT)}

In contrast to CT-TOT, the CF-TOT is defined as a measurement of time-over-threshold at a fixed {\it percentage} or a {\it fraction} of the pulse peak amplitude. Thus, the threshold value varies with pulse amplitude event-by-event, and the same component of the pulse is always sampled. By definition, the CF-TOT is independent of the pulse amplitude and varies only with the pulse shape. Similar to CT-TOT, the fraction should be selected such that at a given threshold one samples mainly the intermediate and slow pulse components, in the range of 1-5$\%$ of peak amplitude.

\section{Experimental configurations}

The following three detector configurations were studied in the context of their PSD performance:
\begin{enumerate}
\item Liquid scintillator coupled to a PMT -- (LS+PMT)
\item Stilbene crystal coupled to a PMT -- (Stilbene+PMT) 
\item Stilbene crystal coupled to a silicon photomultiplier -- (Stilbene+SiPM)
\end{enumerate}

\noindent The liquid scintillator used in the experiment was a self-made NE213-type, encapsulated in a 25 mm diameter Pyrex vial, wrapped with a 3M Viquiti reflector. The stilbene crystal was a 20$\times$20$\times$20 mm$^{3}$ Scintinel detector produced by Inradoptics, wrapped with a Teflon tape. The scintillators were coupled with optical grease to either a 2” PMT (Hamamatsu H1949 assembly with an R-1828-01 PMT) or a quad SiPM (Hamamatsu Quad VUV4 MPPC, model S13371-6050CQ-02). Each SiPM segment has an area of 6$\times$6 mm$^{2}$, with a 0.5 mm gap between segments; it has 13,923 pixels per segment and a geometrical fill-factor of $\sim$ 60$\%$. The window is made of quartz and the pixel pitch is 50 micron~\cite{Hamamatsu1,Hamamatsu2}. The SiPM operation voltage was maintained at -57 V.
The detector signals were digitized using either a Lecroy Waverunner 610ZI oscilloscope with sampling rate of 20 Gs/s or a Tektronix MSO5204B oscilloscope with a sampling rate of 10 Gs/s and the evaluation of the PSD methods was performed off-line. In case of the quad SiPM, the pulses from each segment were summed to get the total pulse. The systems were energy calibrated with $^{137}$Cs and $^{60}$Co gamma-ray sources. All results are expressed in electron-equivalent energy (keVee). The neutrons were produced by a 100 mCi AmBe source, shielded with a 5 mm thick lead shield enclosure to remove the intense 59.54 keV gamma-ray.

\subsection{Classification of neutron/gamma pulses using the Charge Comparison method}

In order to make a comparison between the two TOT methods, we must use two separate sets of gamma-ray and fast neutron pulses. Clean gamma-ray pulses can be collected using various gamma-ray sources; however, neutrons are always accompanied by gamma-rays. Time-of-flight method is probably the best way to generate such separate sets of pulses, however this method was not available to us.
As an alternative, we first classified the pulses obtained from the AmBe source as gamma-rays or neutrons using the standard charge comparison (CC) method. The method uses the ratio of charge in the tail of the pulse, {\it Q$_{tail}$}, to the total charge {\it Q$_{tot}$} as the PSD parameter, as shown in Eq. \ref{eq:y:1}

%
\begin{equation}
\label{eq:y:1}
PSD = Q_{tail}/Q_{tot} = (Q_{tot}-Q_{short})/Q_{tot}
\end{equation}

\begin{figure}[!ht]
\centering
\includegraphics[width=.5\textwidth]{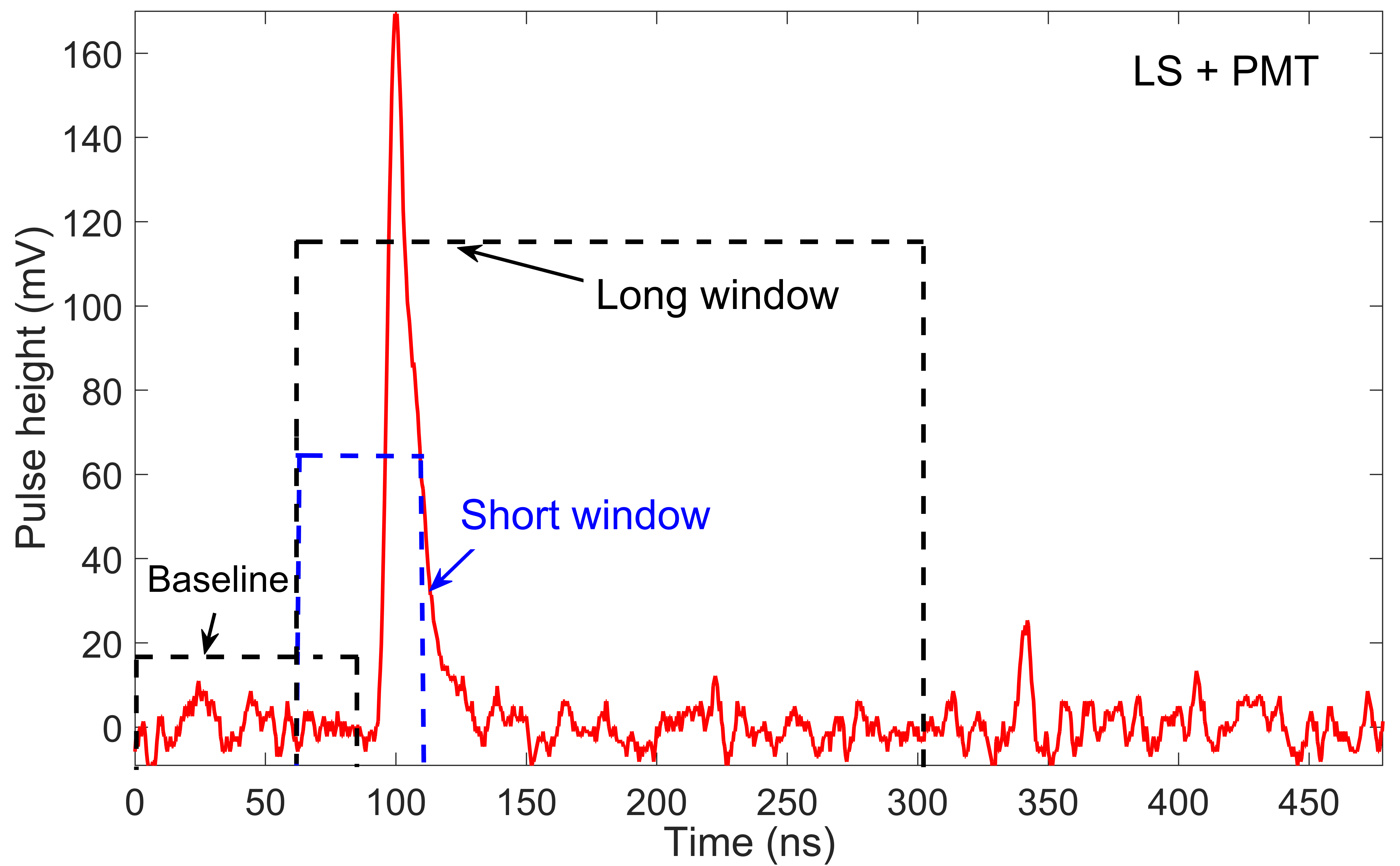}
\includegraphics[width=.5\textwidth]{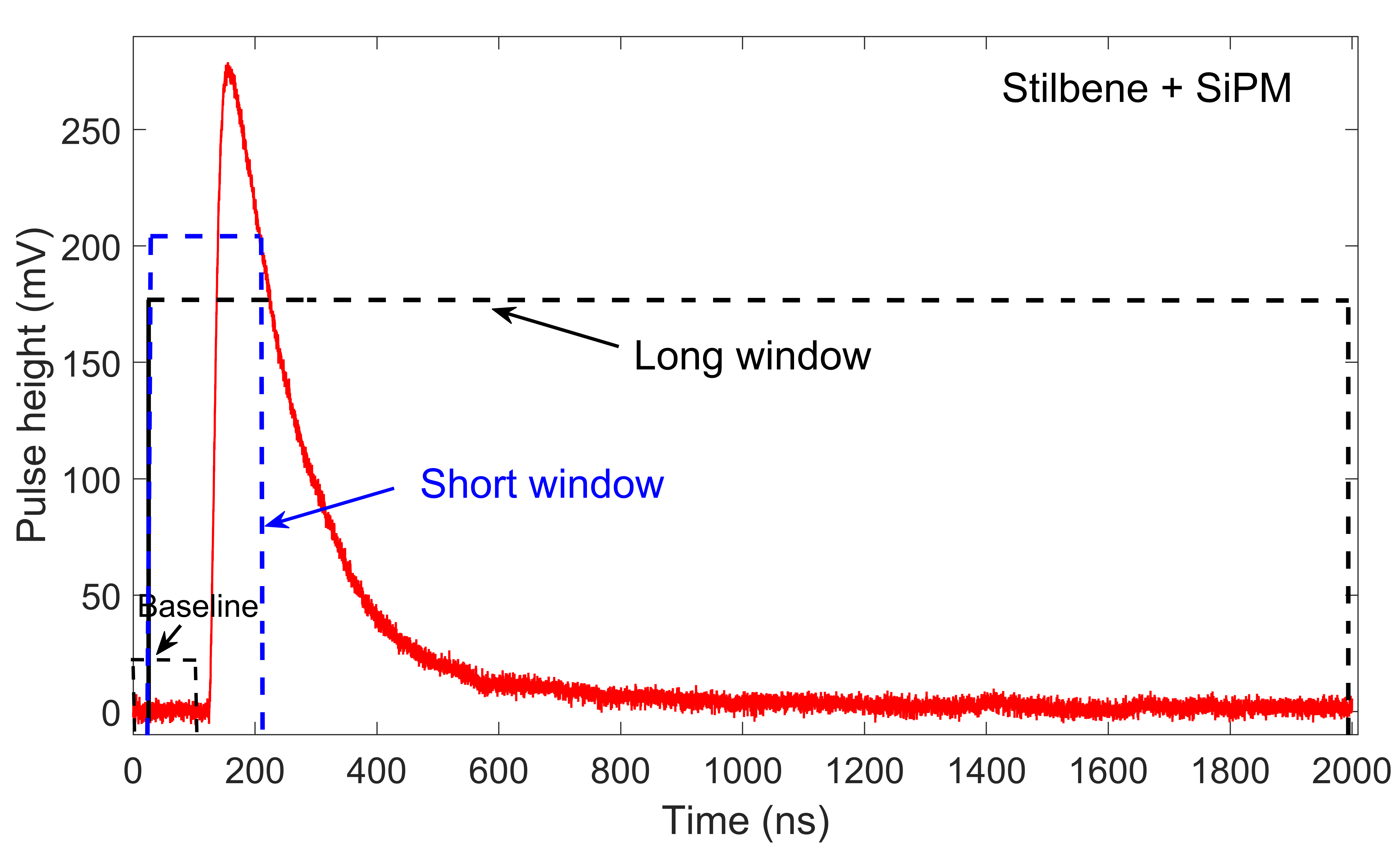}
\caption{\label{fig:iv} Charge integration windows for LS+PMT (top) and Stilbene+SiPM (bottom).}
\end{figure}

{\noindent}where {\it Q$_{short}$} is defined as the integral charge in the short window used to determine the prompt light emission. The total charge, {\it Q$_{tot}$}, is evaluated by integrating over the long window. It is critical to optimize the integration windows to obtain the best possible PSD. The optimized widths of the specific integration time windows used in this analysis for the three detector configurations are provided in Table \ref{tab:table1}. The time windows for the baseline-subtracted pulses start 40 ns and 120 ns before the PMT and the SiPM peak amplitudes, respectively. Approximately 10,000 pulses were collected for each configuration. The energy threshold was set at 100 keVee. \\
\noindent An illustration of the charge integration windows for the PMT and SiPM detector configurations are shown in Figure \ref{fig:iv}. 

\begin{table}[!ht]
\caption{Optimized integration time windows for the three detector configurations}
\label{tab:table1}
\centering
\begin{tabular}{|c|c|c|}
\hline
\textbf{Configuration} & \textbf{Short window (ns)} & \textbf{Long window(ns)} \\ \hline
LS+PMT                 & 50 ns                      & 240 ns                    \\ \hline
Stilbene+PMT           & 50 ns                      & 280 ns                    \\ \hline
Stilbene+SiPM          & 200 ns                     & 2200 ns                   \\ \hline
\end{tabular}
\end{table}

\begin{figure}[!ht]
\centering
\includegraphics[width=.5\textwidth]{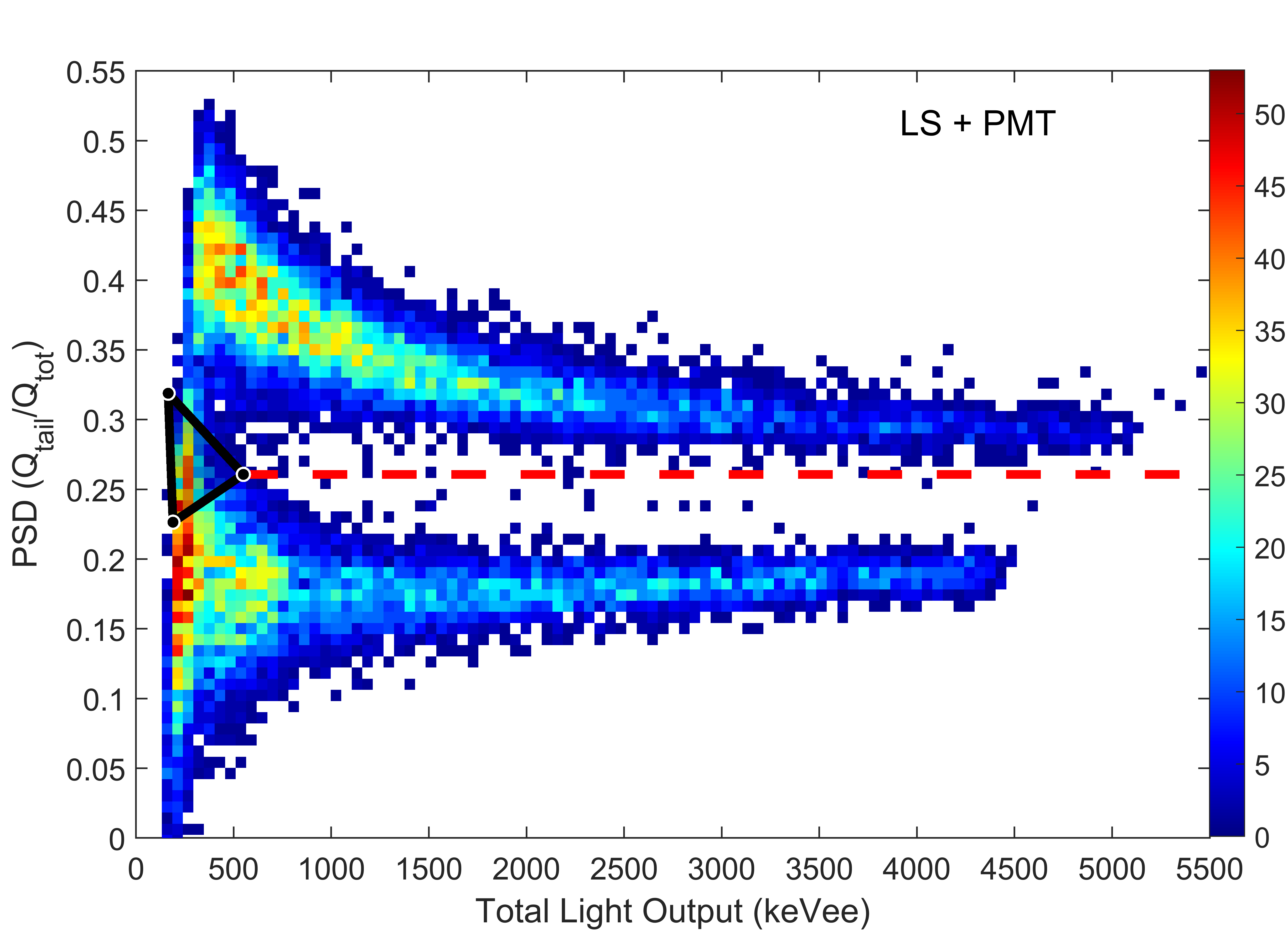}
\includegraphics[width=.5\textwidth]{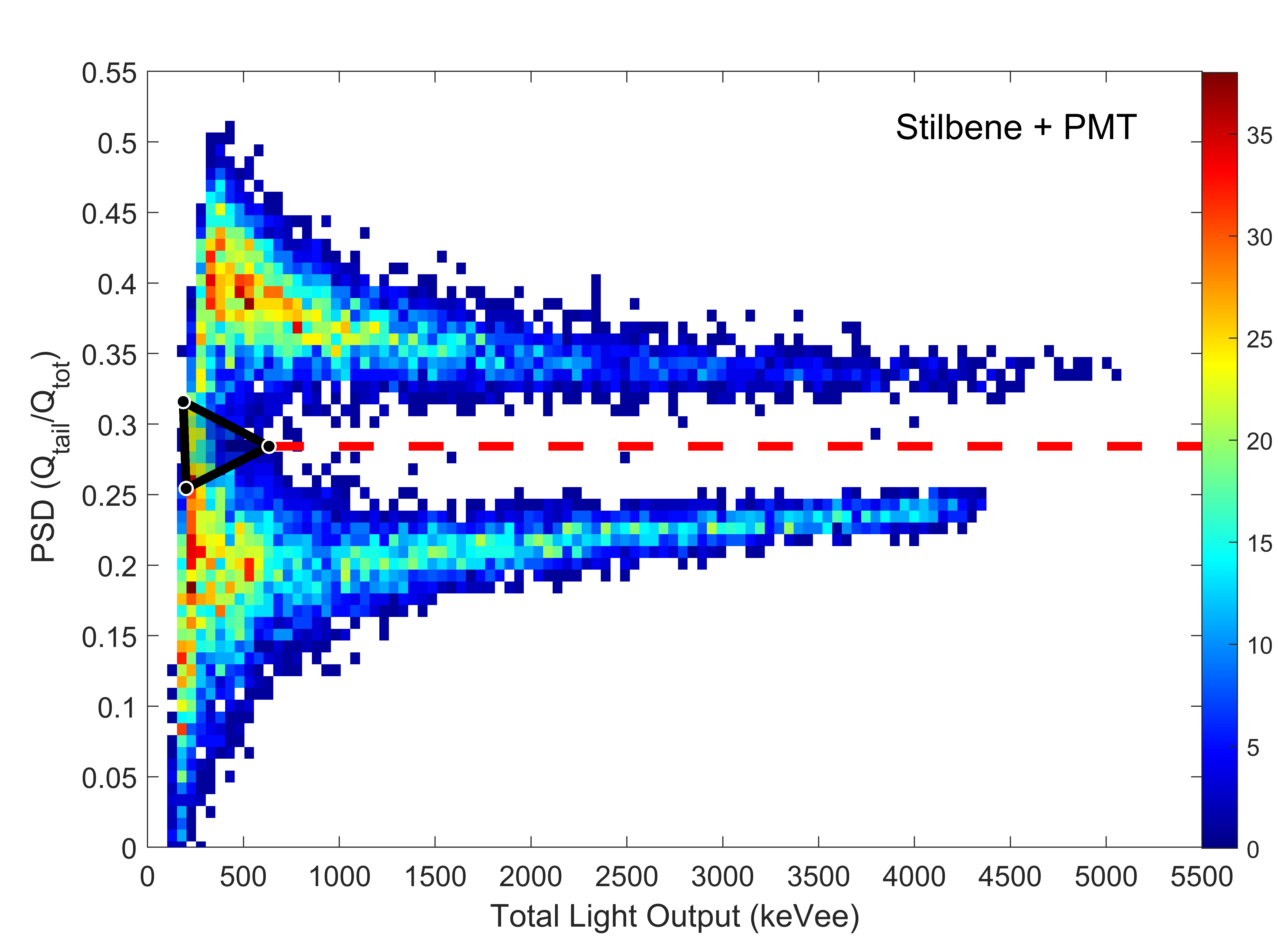}
\includegraphics[width=.5\textwidth]{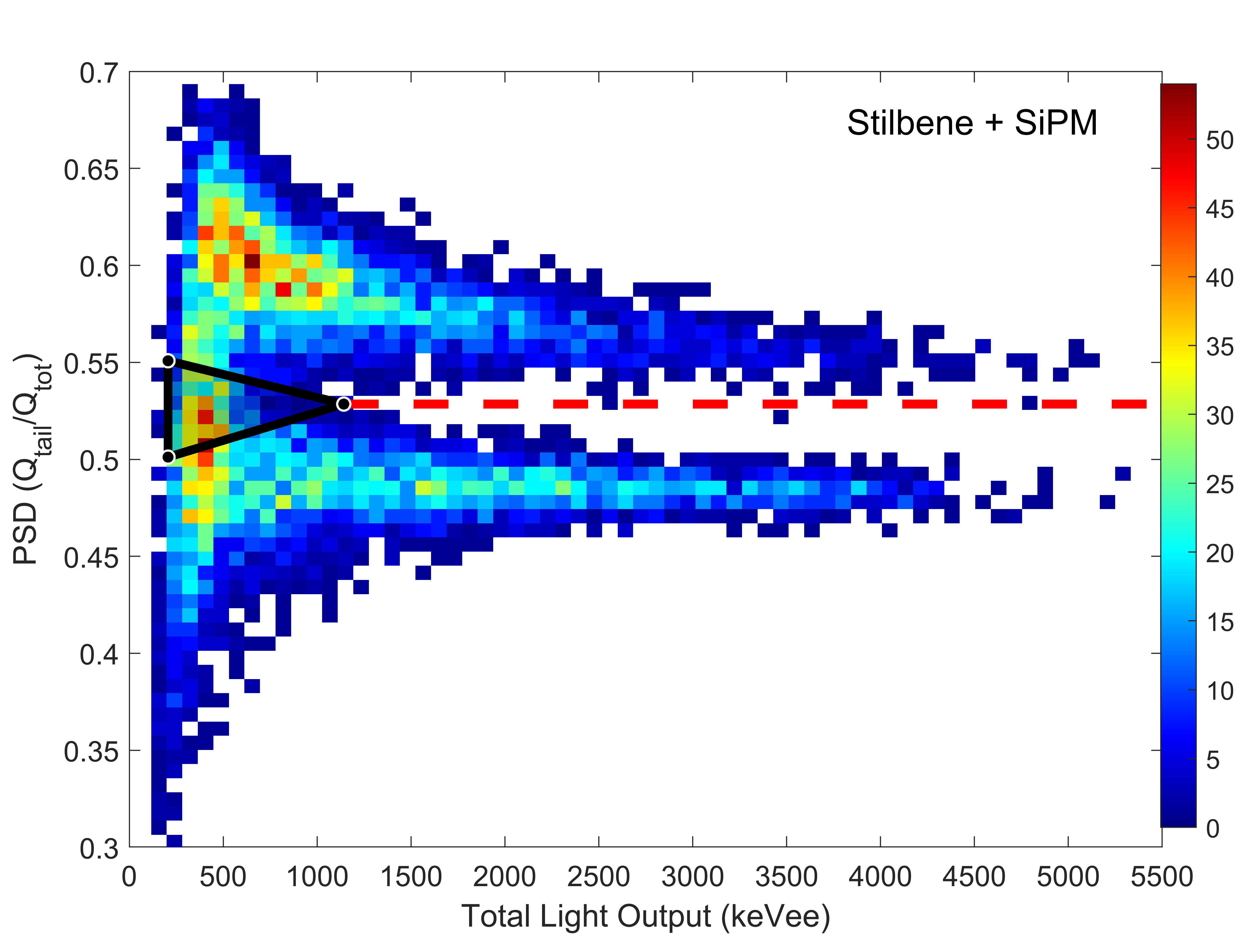}
\caption{\label{fig:v} Charge comparison PSD vs. particle total light output in keVee, top - LS+PMT, center - Stilbene+PMT, bottom - Stilbene+SiPM. The black triangular cut between the two regions represents a zone of ambiguity and is removed from the dataset offline to perform a more reliable neutron and gamma-ray classification (see text below). The fixed PSD threshold shown with the red dashed line is used to classify the rest of the pulses. The pulses below this PSD threshold were classified as gamma-rays and above it as neutrons.}
\end{figure}

\begin{figure}[!ht]
\centering
\includegraphics[width=.45\textwidth]{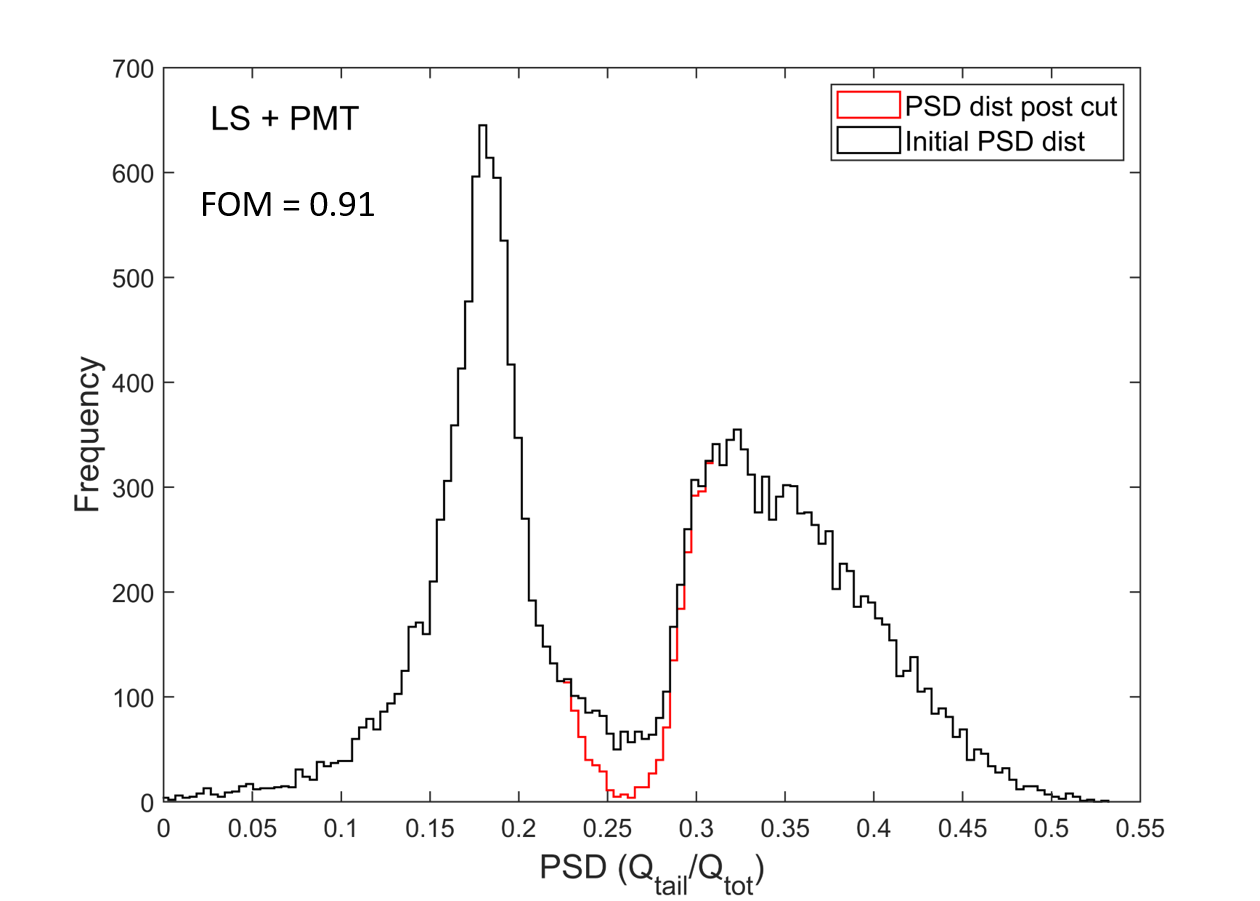}\\
\includegraphics[width=.45\textwidth]{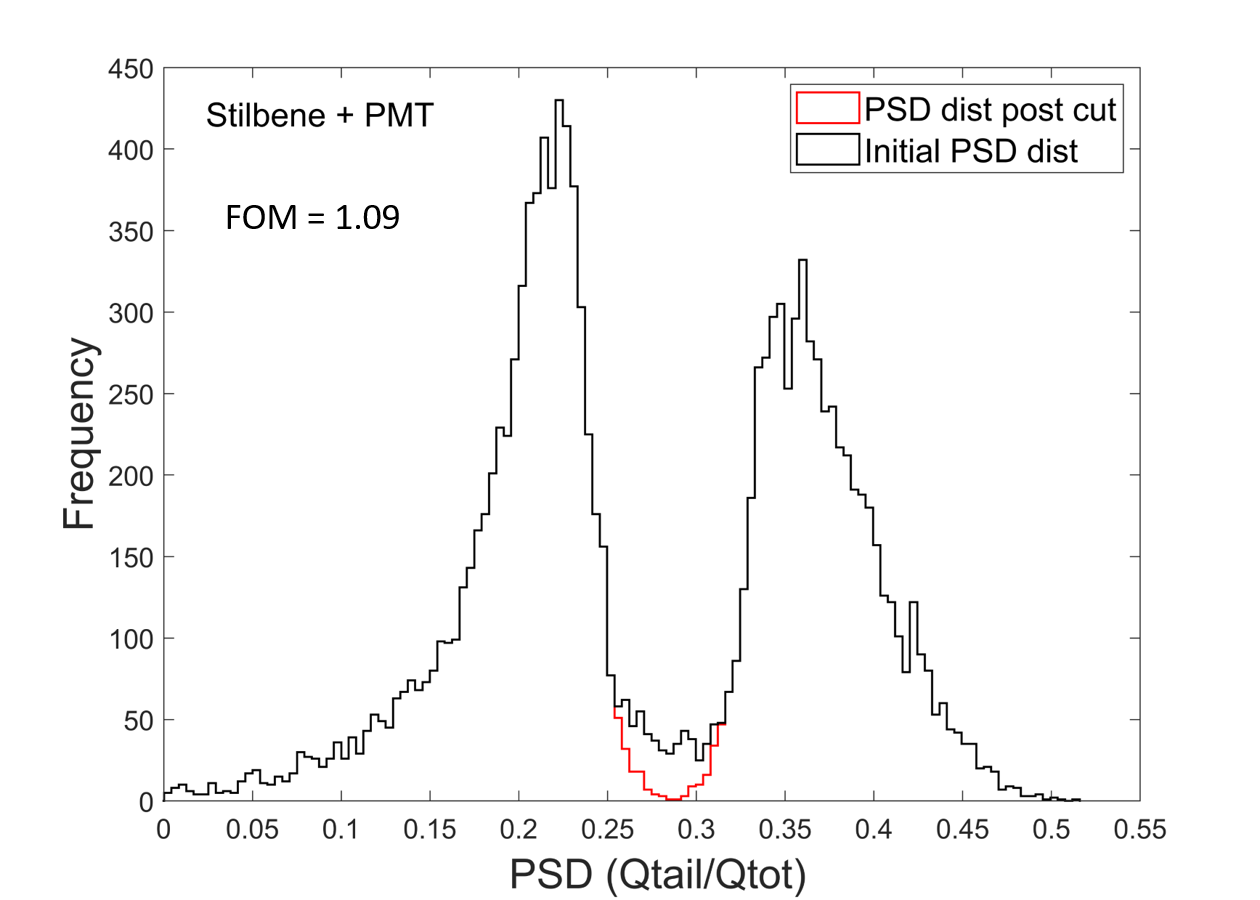}\\
\includegraphics[width=.45\textwidth]{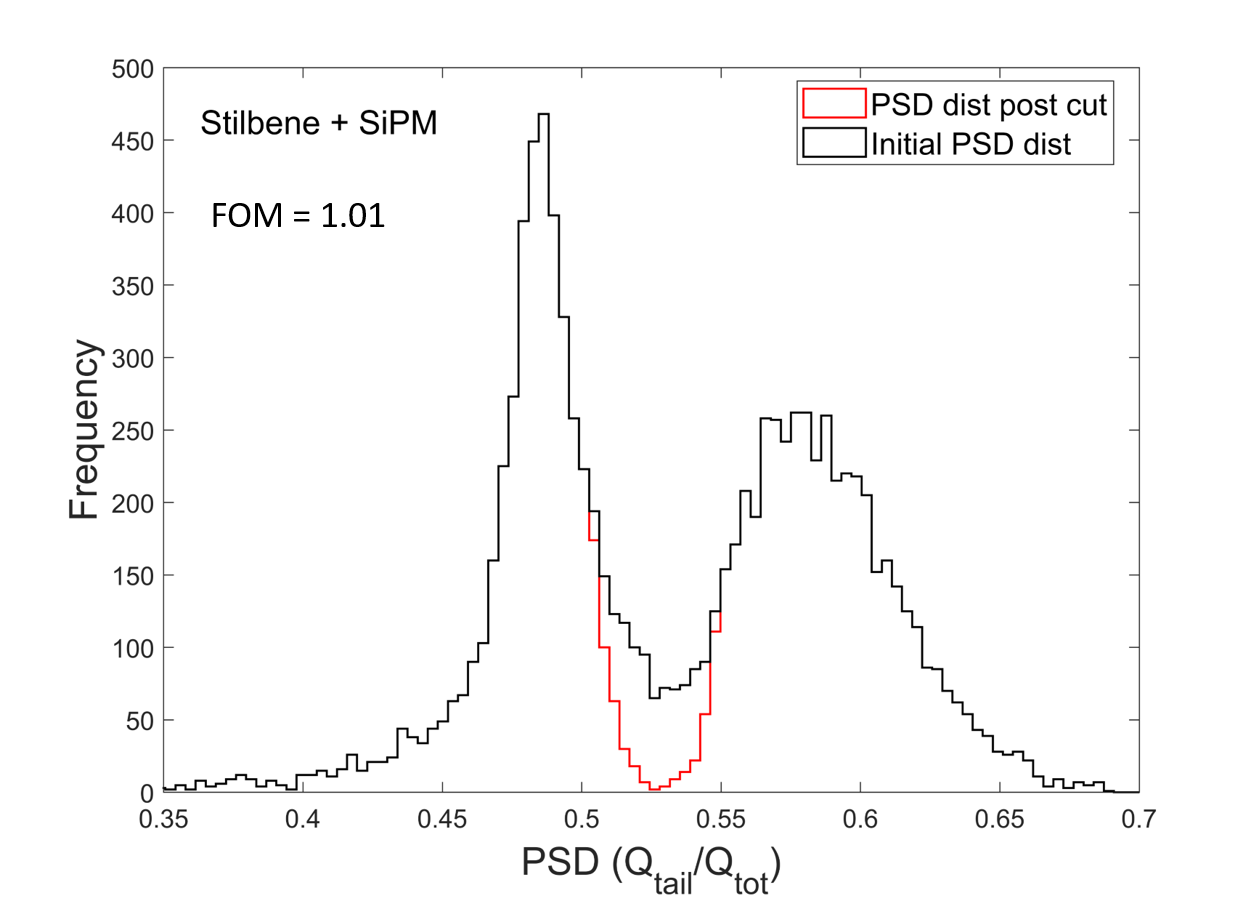}
\caption{\label{fig:vi} PSD frequency distributions for the LS+PMT, Stilbene+PSD and Stilbene+SiPM configurations before and after the removal of the ambiguous region.}
\end{figure}

\noindent Figure \ref{fig:v} shows the CC PSD value as a function of the total event light output expressed in keVee, for the three configurations. 
As can be seen clearly in Figure \ref{fig:v}, at low particle energies, the upper tail of the gamma-ray distribution merges with the lower tail of the neutron region, creating an ambiguity in pulse classification. Since our goal is to classify the events with the highest possible certainty, we removed 
a small fraction of the events in the valley region -- the triangular region shown in black -- between the gamma-ray and neutron distributions,
at the cost of a $\sim$5$\%$ loss in the total number of detected events. After the removal of this region, we fixed a PSD threshold to classify the rest of the pulses. The pulses below this PSD threshold were classified as gamma-rays and above it as neutrons. The dashed red line in Figure \ref{fig:v} helps to guide the eye. The PSD thresholds used for the three configurations are 0.26, 0.28 and 0.53, respectively. The pulses were then color-coded for all consequent studies and assigned red color for gamma-rays and black for neutrons.\\
\noindent Next, we present the above data as frequency distributions in order to calculate the standard figure of merit, (FOM). FOM is defined as the ratio of the distance between PSD frequency distribution neutron and gamma-ray peak positions to the sum of their full widths at half-maximum. The calculated FOM values were 0.91, 1.09 and 1.01 for the LS+PMT, Stilbene+PMT and Stilbene+SiPM configurations respectively. Figure \ref{fig:vi} shows a comparison between the PSD frequency distributions before and after the removal of the triangular region, resulting in an improvement in the neutron/gamma-ray classification.

\section{Analysis of the TOT methods}
\noindent After the classification of the pulses as neutrons or gamma-rays, we performed the TOT analysis for each configuration.

\subsection{Constant threshold TOT (CT-TOT)}

\begin{figure}[!ht]
\centering
\includegraphics[width=.45\textwidth]{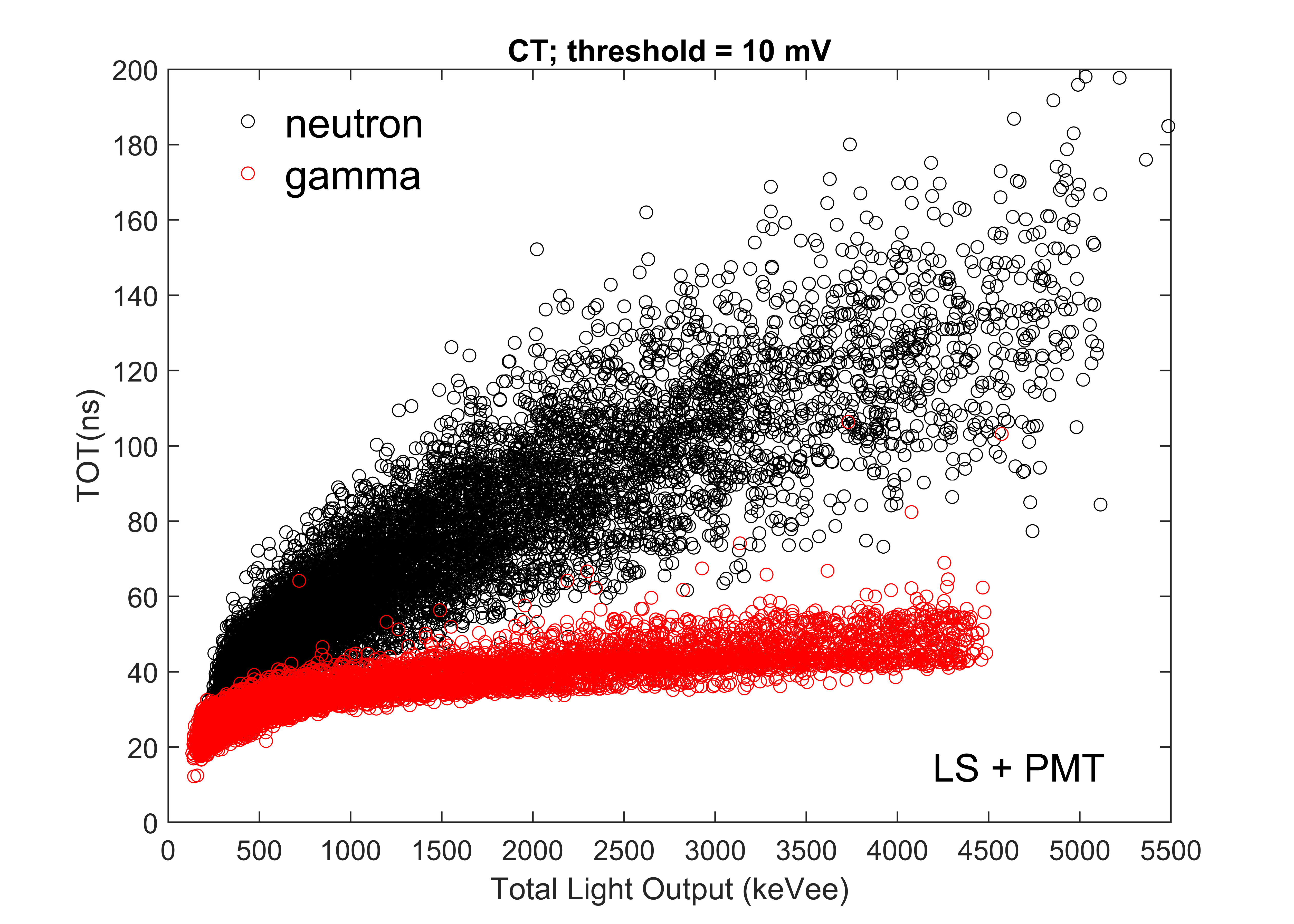}\\
\includegraphics[width=.45\textwidth]{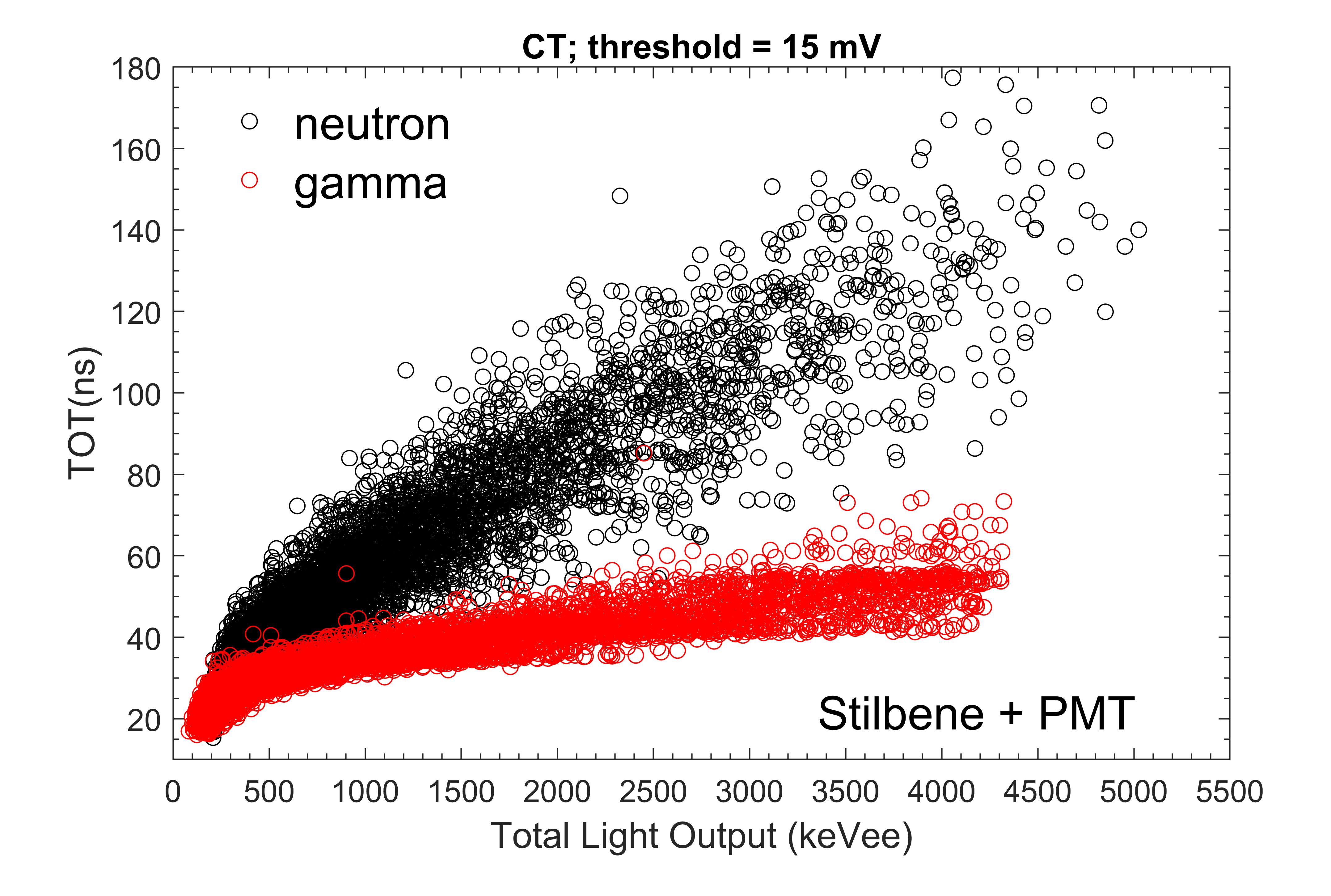}\\
\includegraphics[width=.45\textwidth]{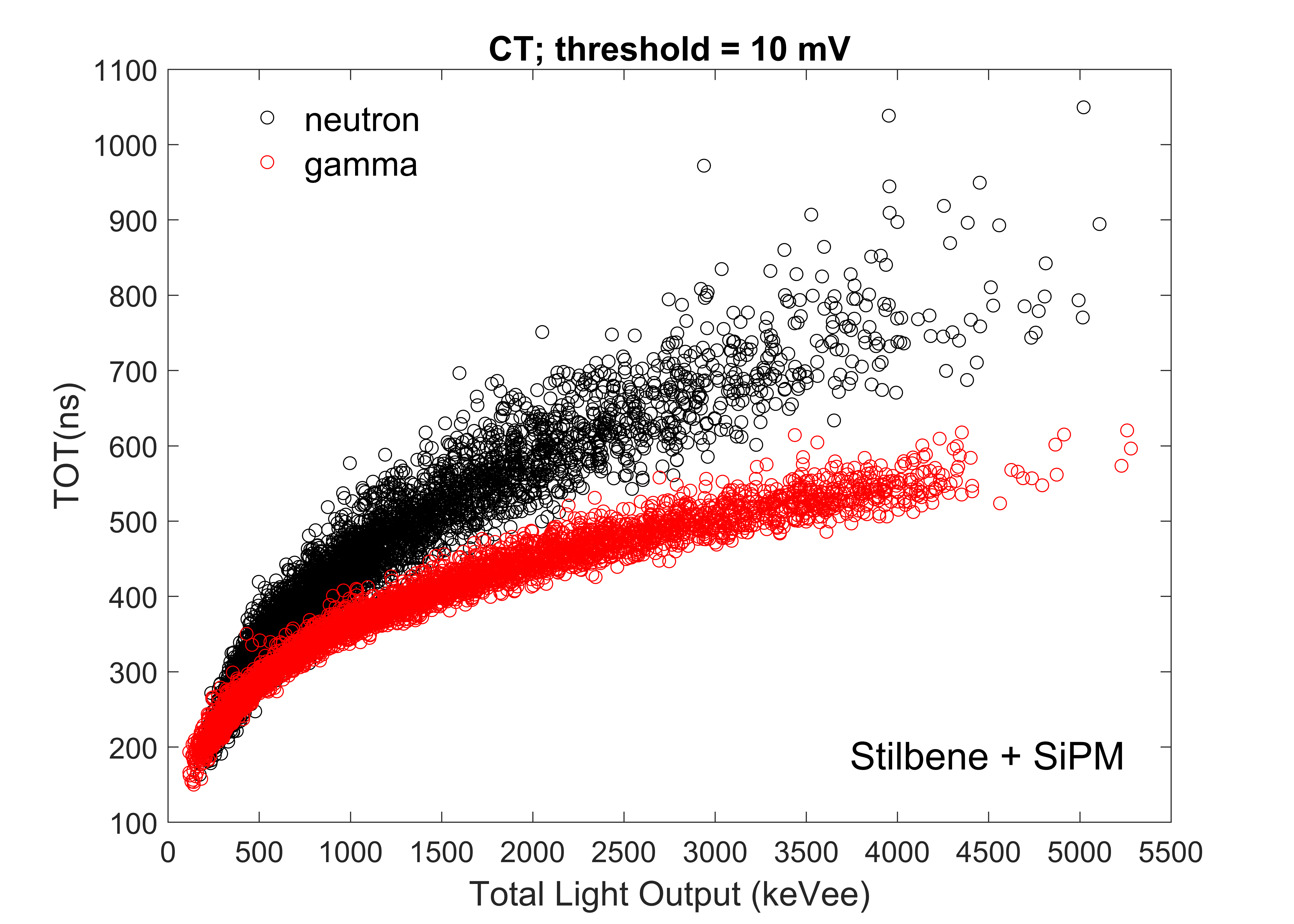}
\caption{\label{fig:vii} CT-TOT vs. total light output in keVee; top - LS+PMT, center - Stilbene+PMT, bottom 
Stilbene+SiPM. The amplitude threshold levels are indicated above the figures.}
\end{figure}

\noindent Figure \ref{fig:vii} shows CT-TOT for the three detector configurations. For each case, the threshold voltages were chosen to maximize the separation and to minimize the broadening of the distribution due to the noise and pulse irregularities. The chosen thresholds were 10 mV, 15 mV and 10 mV for the LS+PMT, Stilbene+PMT and Stilbene+SiPM respectively. As can be observed, the CT-TOT neutron/gamma separation is poor at low energies, because at low pulse amplitudes the threshold voltage samples mainly the short pulse component, which is not very different for the two types of particles. The neutron and gamma bands are indistinguishable up to $\sim$ 1000 keVee, for the CT-TOT analysis.

\subsection{Constant fraction TOT (CF-TOT)}

\begin{figure}[!ht]
\centering
\includegraphics[width=.45\textwidth]{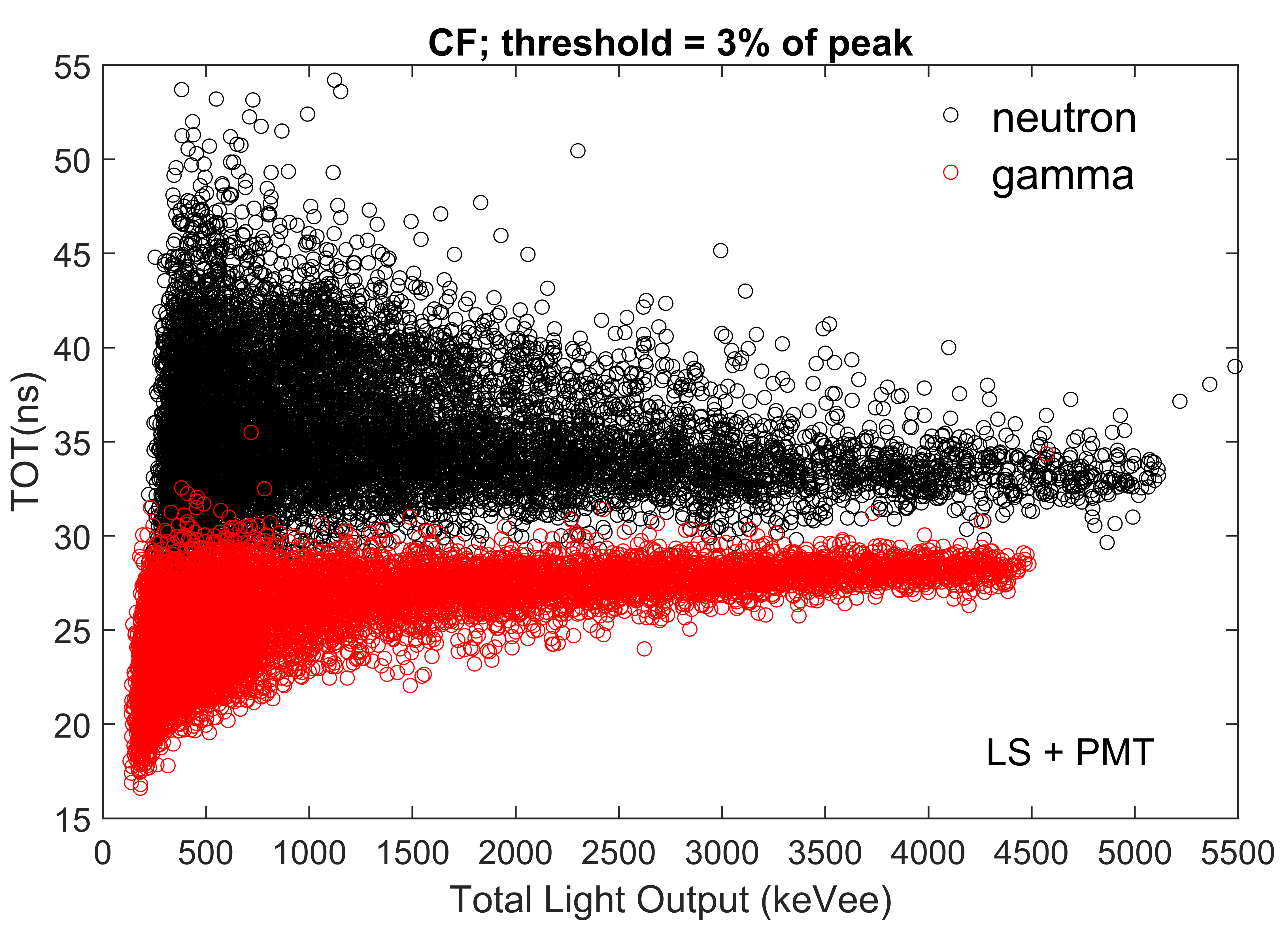}\\
\includegraphics[width=.45\textwidth]{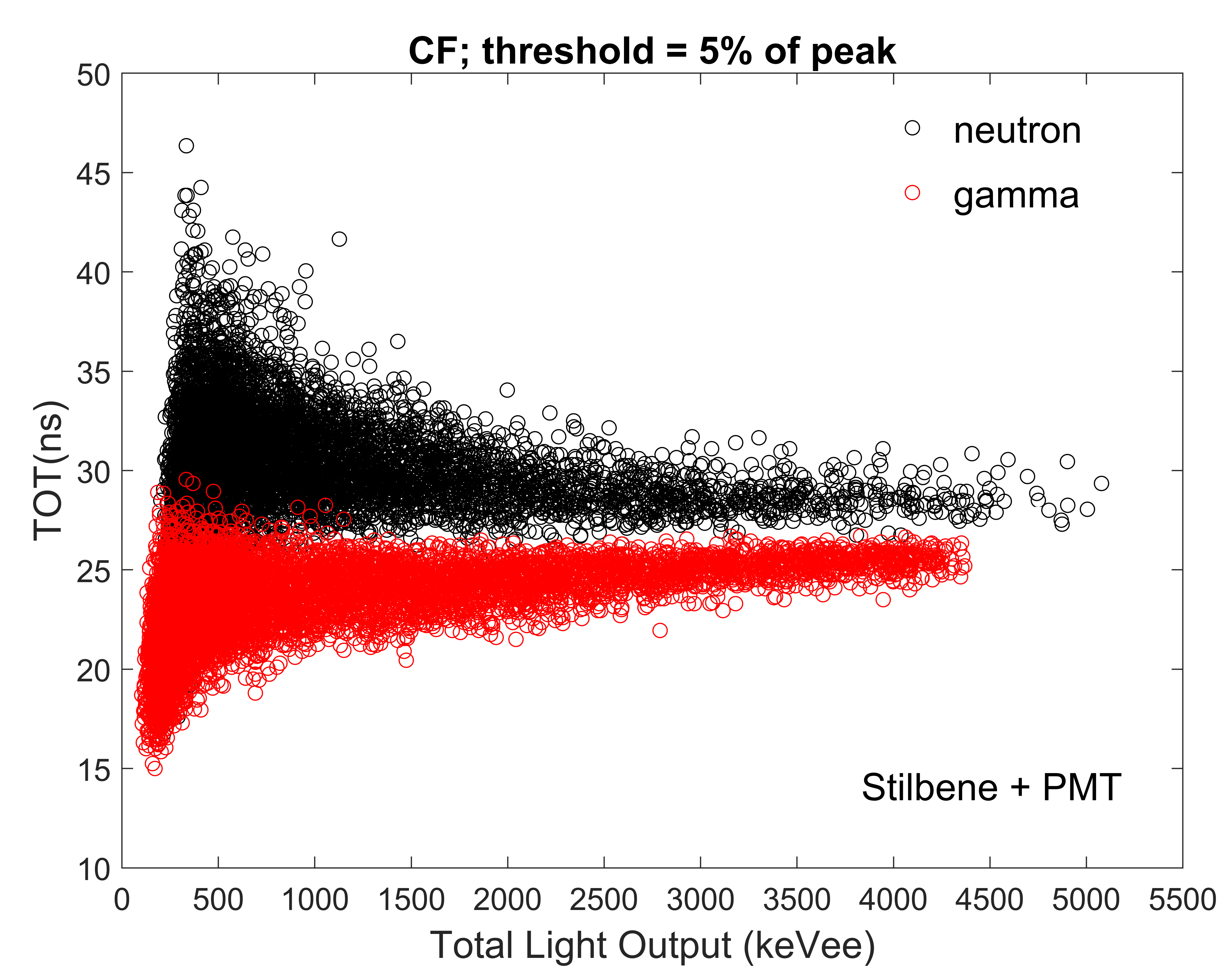}\\
\includegraphics[width=.45\textwidth]{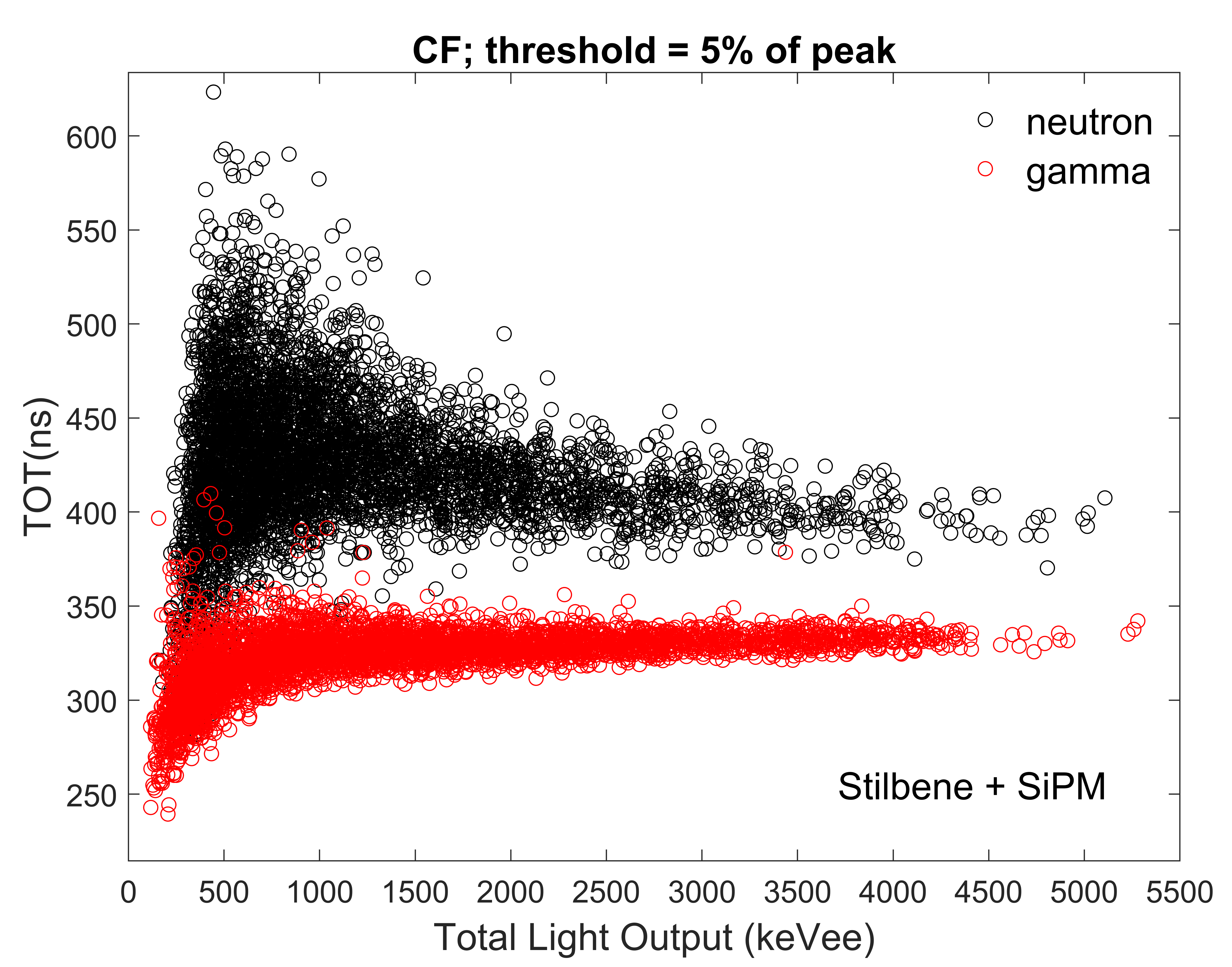}
\caption{\label{fig:viii} CF-TOT vs. total light output in keVee; top - LS+PMT, center - Stilbene+PMT, bottom -
Stilbene+SiPM. The constant fraction threshold levels are indicated above the figures.}
\end{figure}

\noindent Figure \ref{fig:viii} shows CF-TOT as a function of the total light output for the three configurations. As expected, the CF-TOT is independent of particle energy and is broadened mainly by the pulse quality and its smoothness. 

\noindent The natural smoothing of the pulse in the SiPM case due to its slow response is removing the sharp irregularities in the pulse. The selection of the fraction is a compromise between particle separation and noise.
\section{Comparative study of the TOT PSD methods}

\begin{figure}[!ht]
\centering
\includegraphics[width=.45\textwidth]{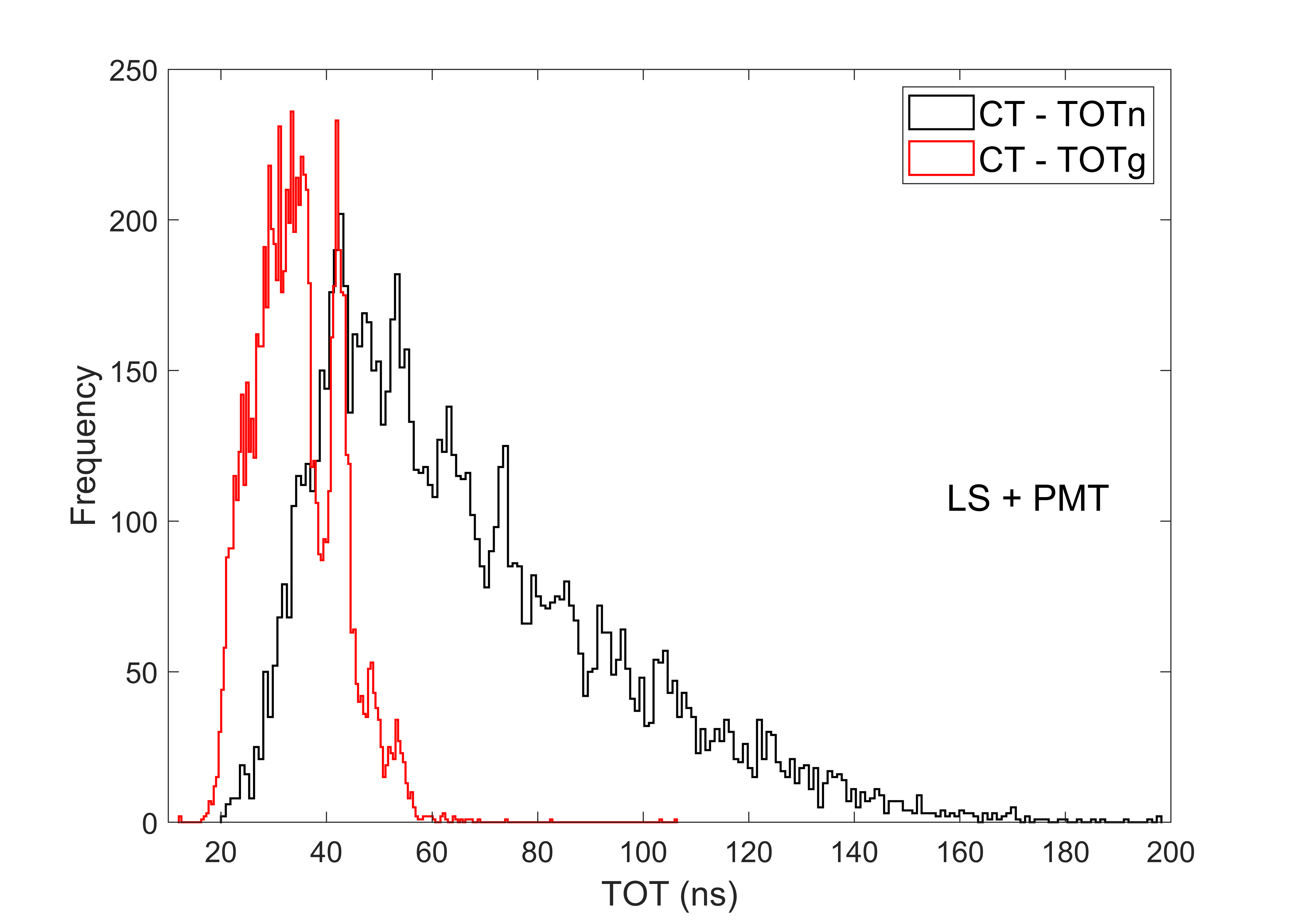}\\
\includegraphics[width=.45\textwidth]{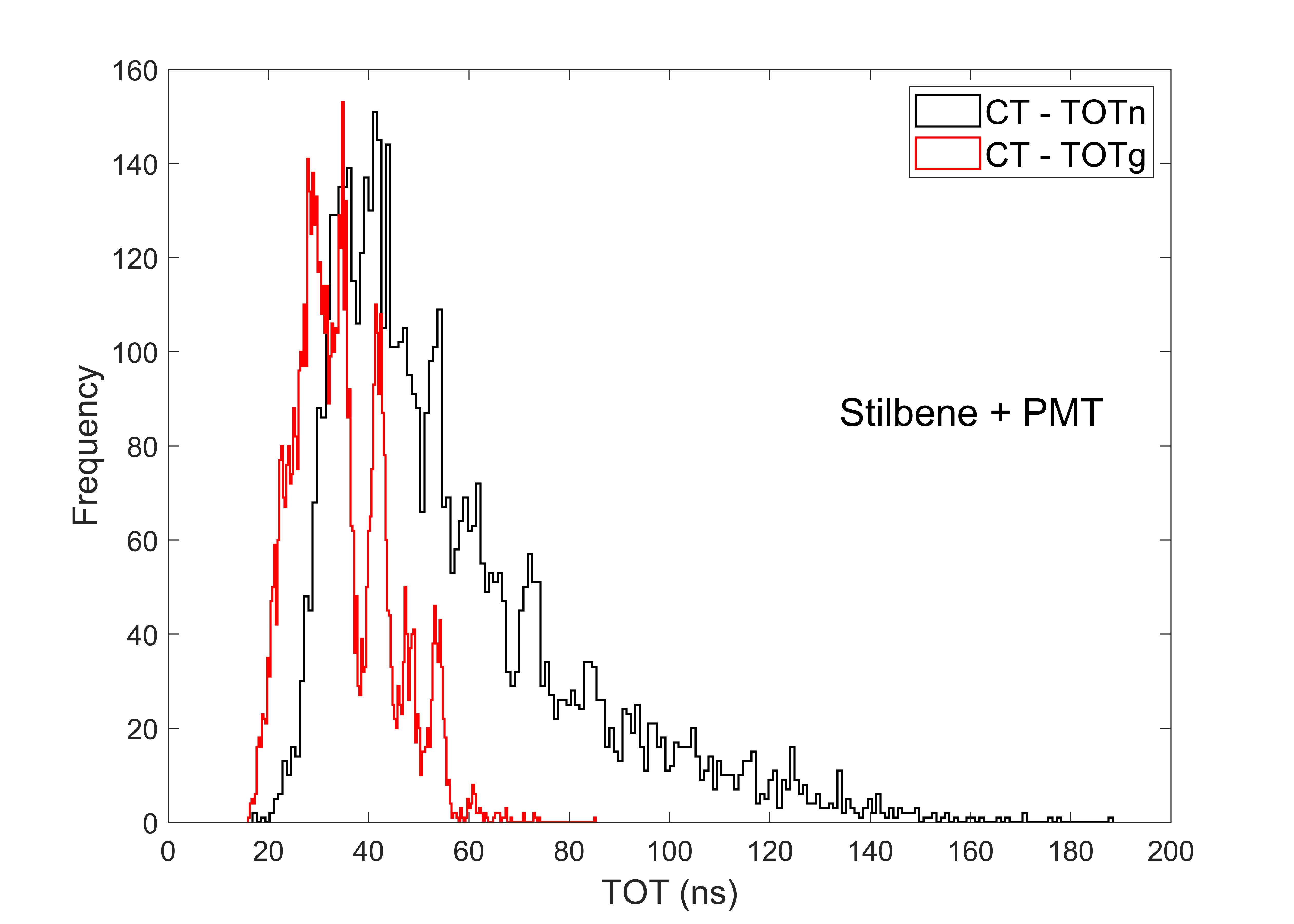}\\
\includegraphics[width=.45\textwidth]{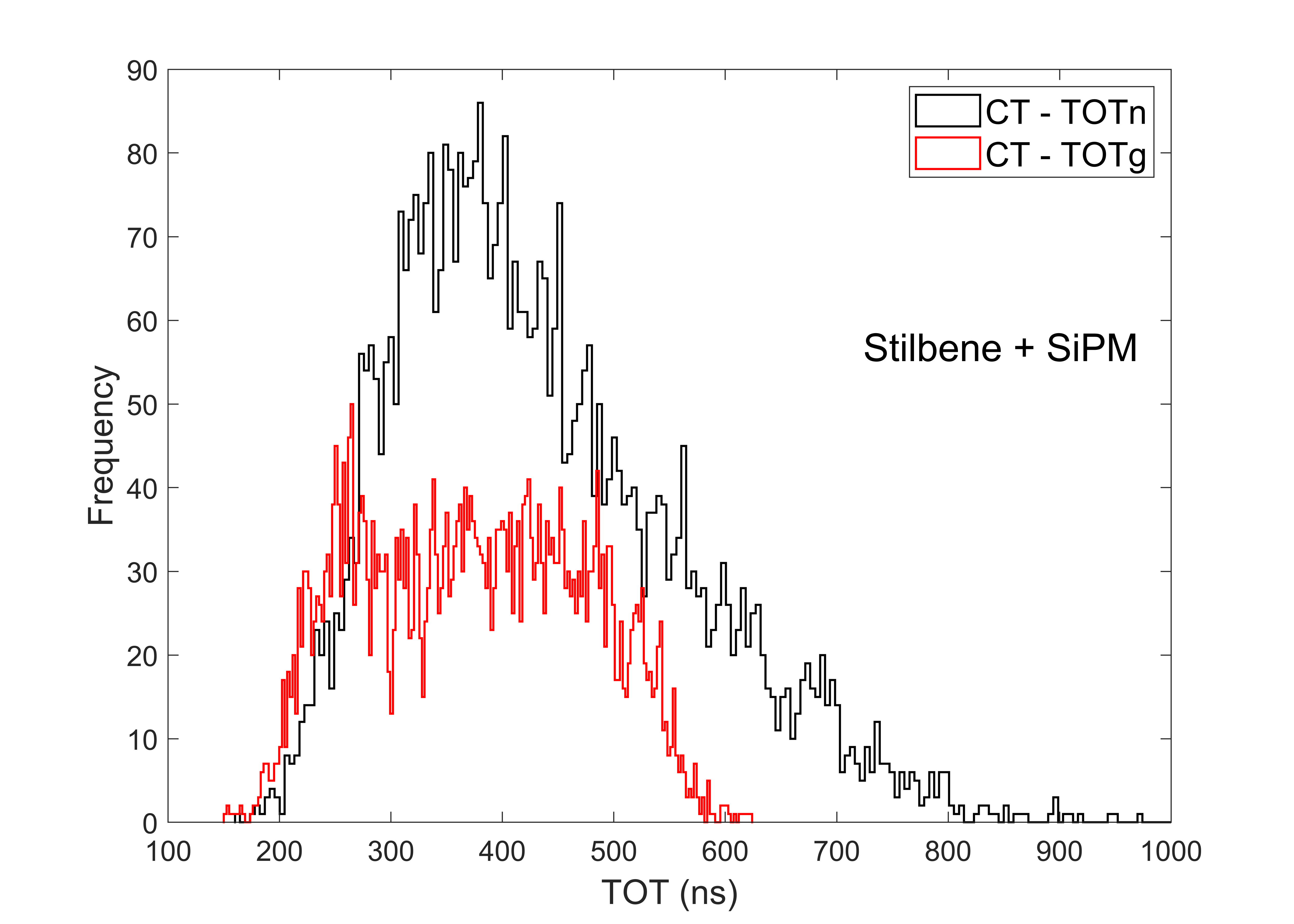}
\caption{\label{fig:ix} Frequency distribution of the CT-TOT method for the three configurations ; LS+PMT - top, Stilbene+PMT - middle, and Stilbene+SiPM - bottom.}
\end{figure}

\noindent It is customary to compare the quality of pulse shape discrimination using the standard FOM method as defined in section 3. We attempted to use this approach for the comparison of the two TOT approaches. Figure \ref{fig:ix} shows the frequency distribution of the CT-TOT method for the three configurations. As expected, the two distributions are not well separated and clearly the standard FOM method cannot be applied for the CT-TOT analysis.

\begin{figure}[!ht]
\centering
\includegraphics[width=.5\textwidth]{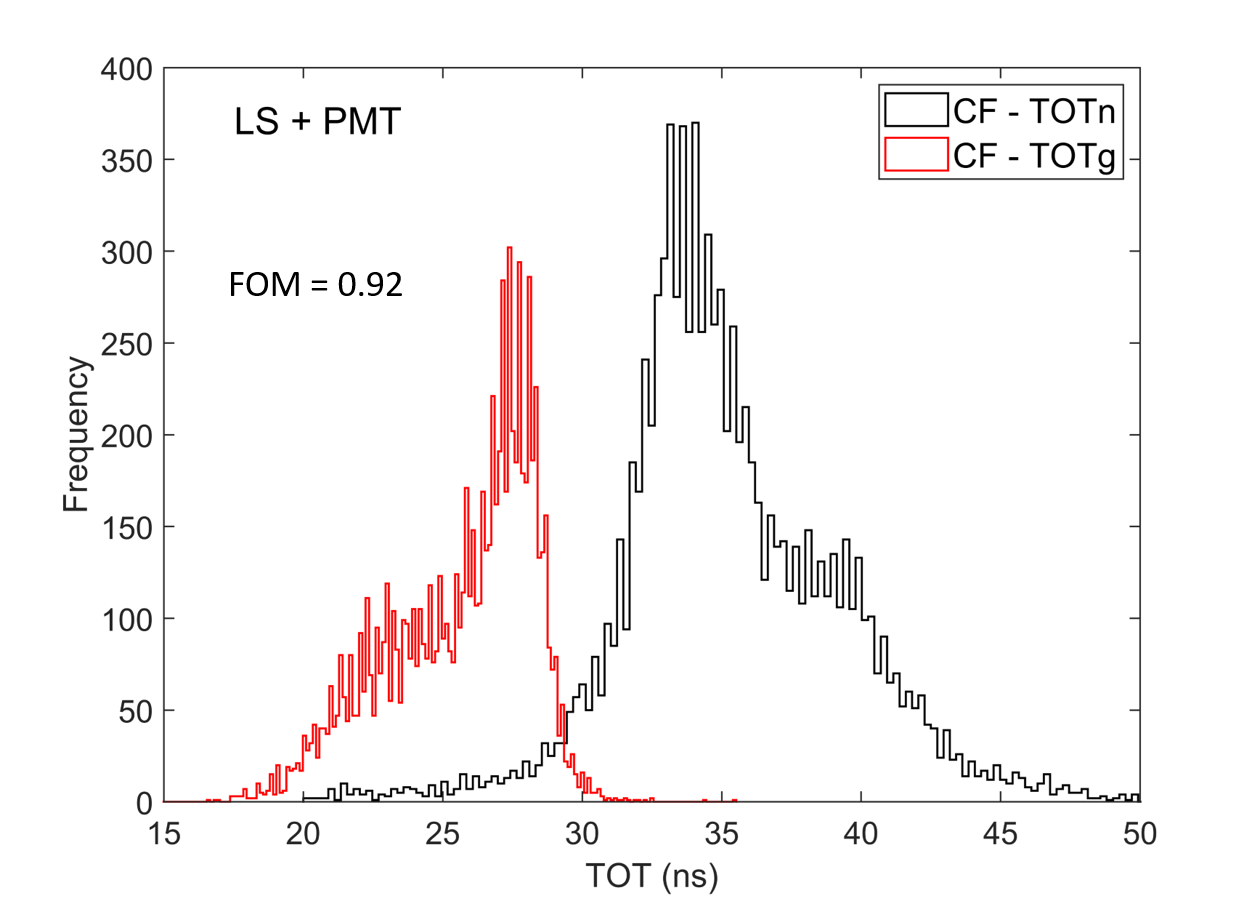}
\includegraphics[width=.5\textwidth]{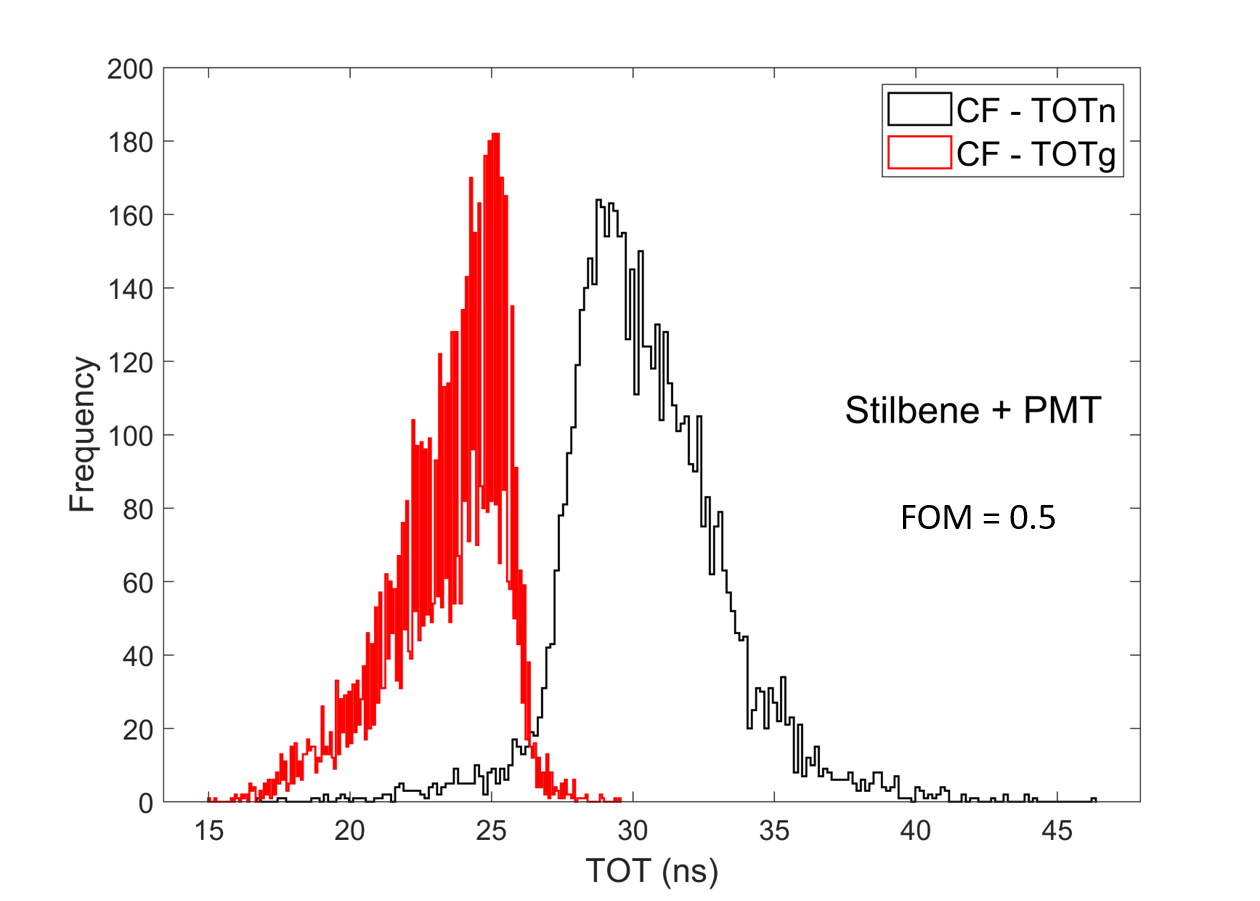}
\includegraphics[width=.5\textwidth]{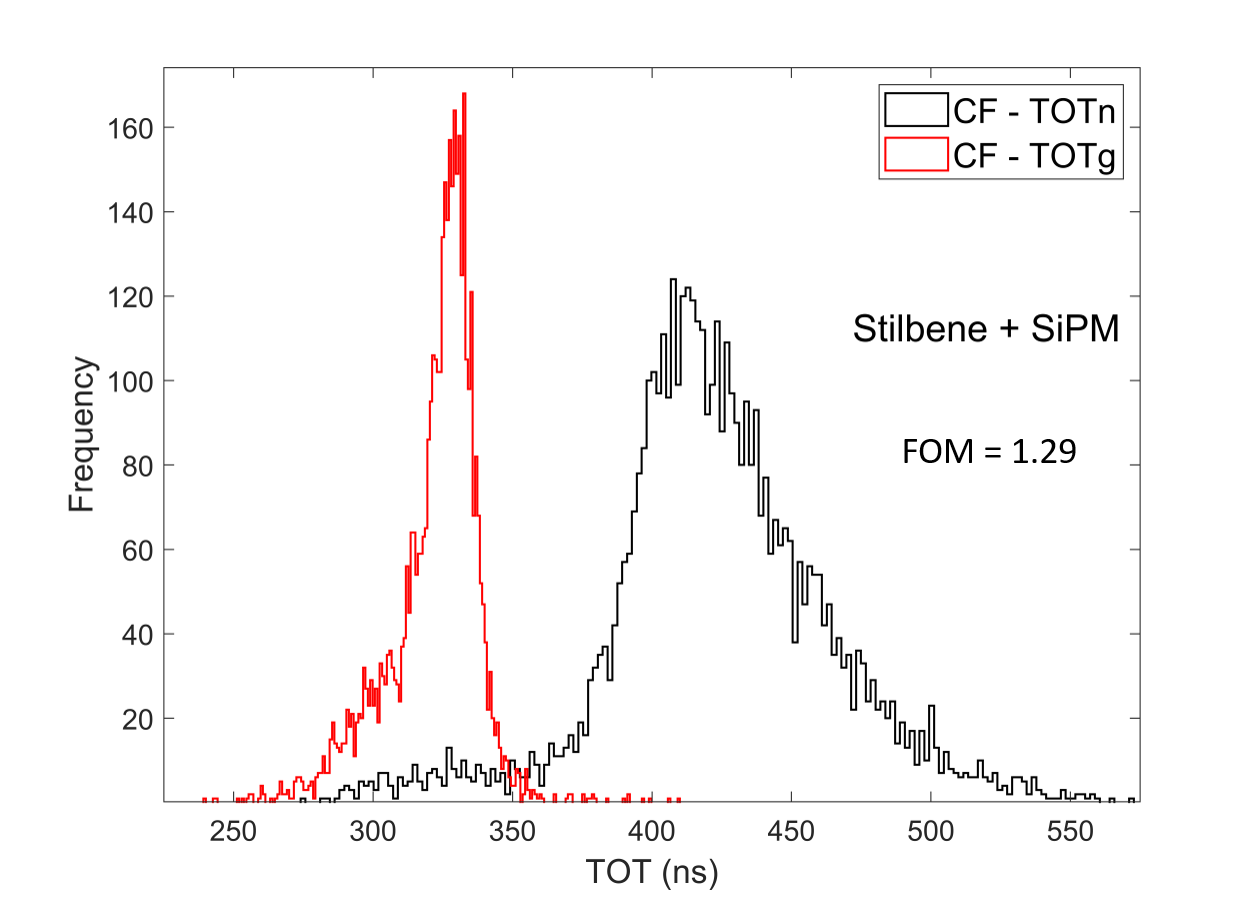}
\caption{\label{fig:x} Frequency distributions of the CF-TOT method for the three configurations ; LS+PMT - top, Stilbene+PMT - middle, and Stilbene+SiPM -- bottom.}
\end{figure}

\noindent The frequency distribution for the CF-TOT method are shown in Figure \ref{fig:x}. Due to the fact that the CF-TOT peaks are quite well resolved, one can calculate here the FOM values. The calculated FOM values for the LS+PMT, Stilbene+PMT and the Stibene+SiPM configurations were 0.92, 0.5 and 1.29, respectively. In general, it appears that for the CF-TOT method, the shapes of the distributions and the FOM values are worse than those obtained by the CC method (see Figure \ref{fig:vi}). Although the FOM value for the Stilbene+SiPM configuration appears to be better than the FOM value for CC (1.01), there is a significant overlap between the gamma-rays and neutrons in the wings of the distributions.\\

\begin{figure}[!ht]
\centering
\includegraphics[width=.5\textwidth]{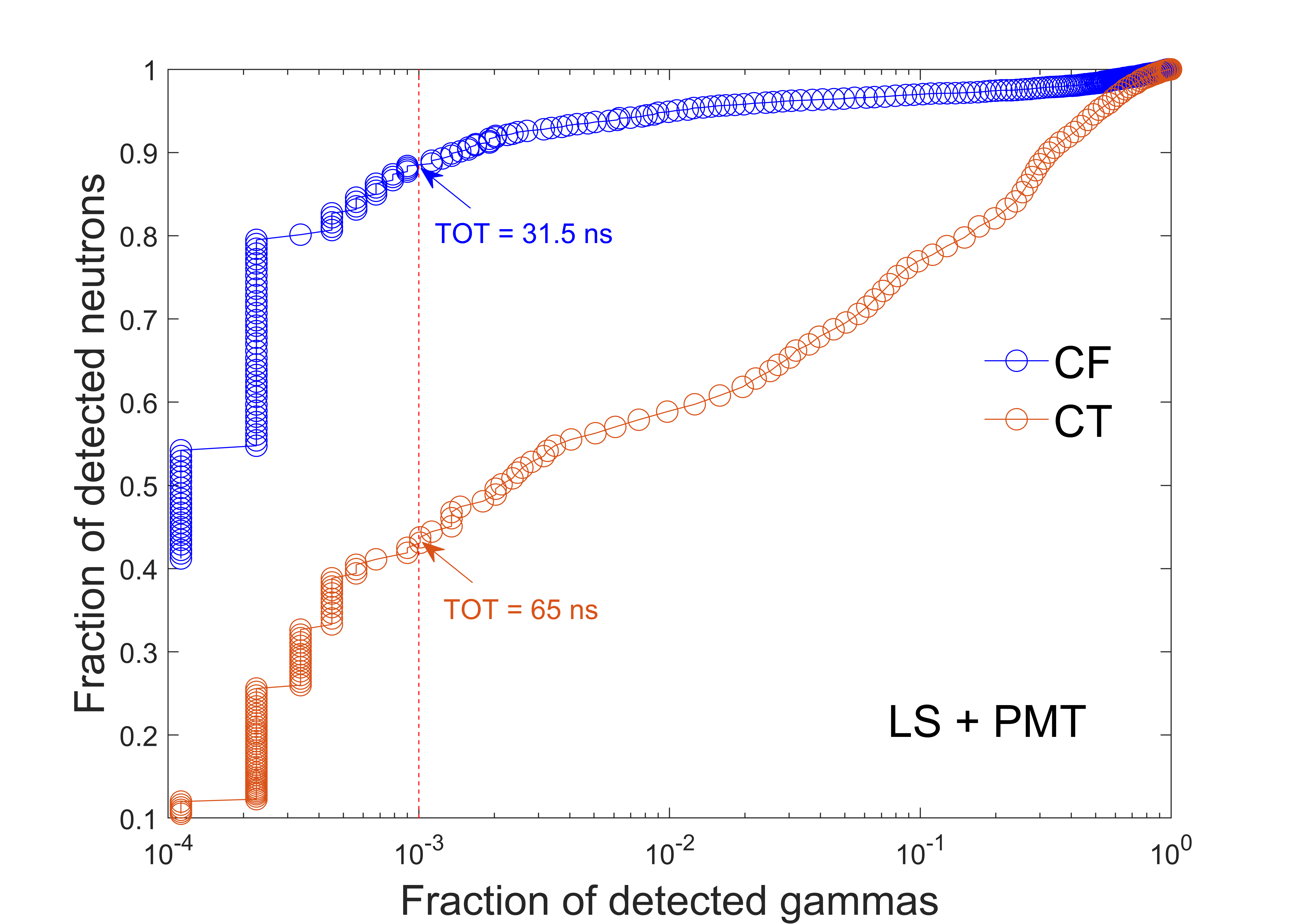}
\includegraphics[width=.5\textwidth]{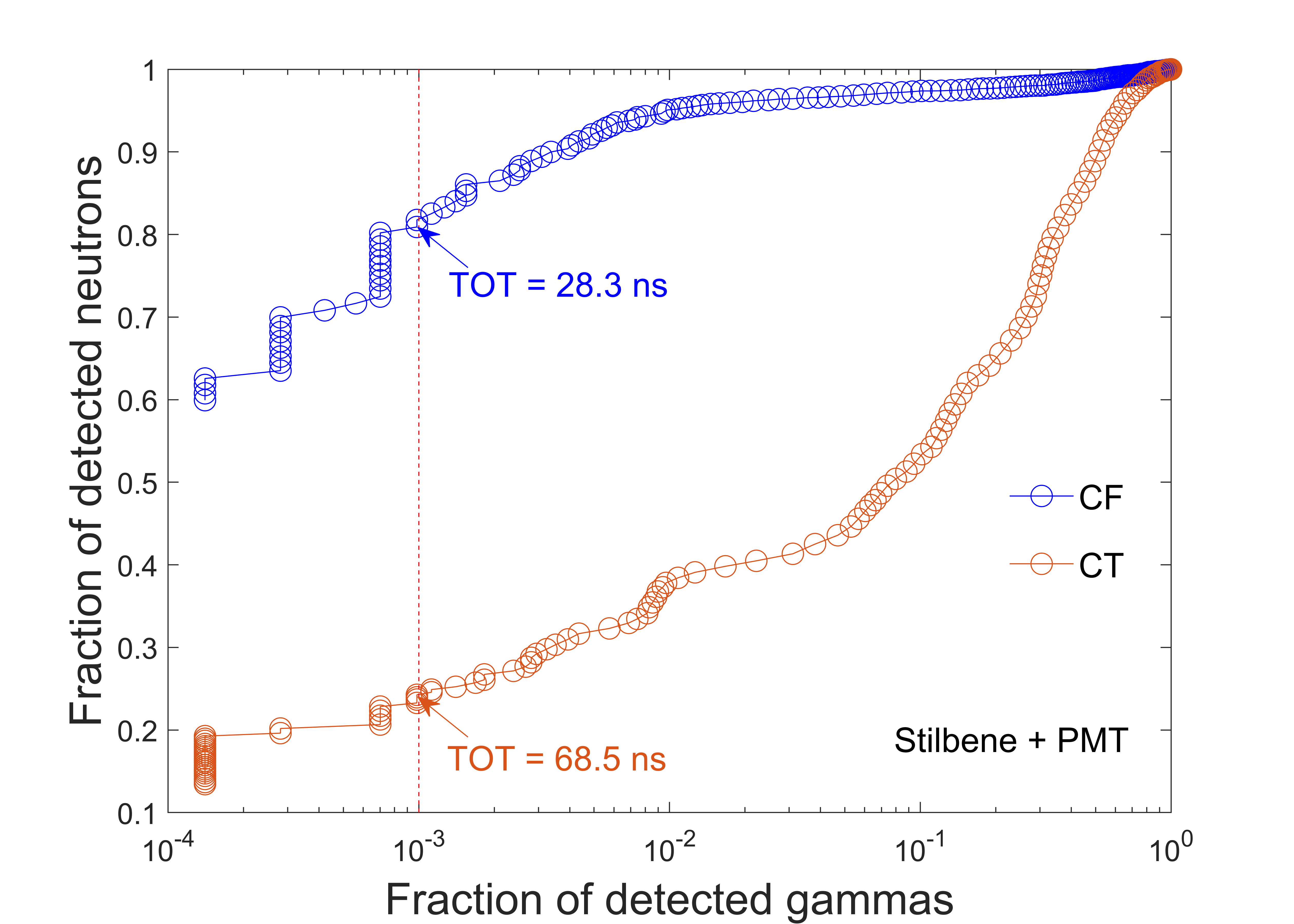}
\includegraphics[width=.5\textwidth]{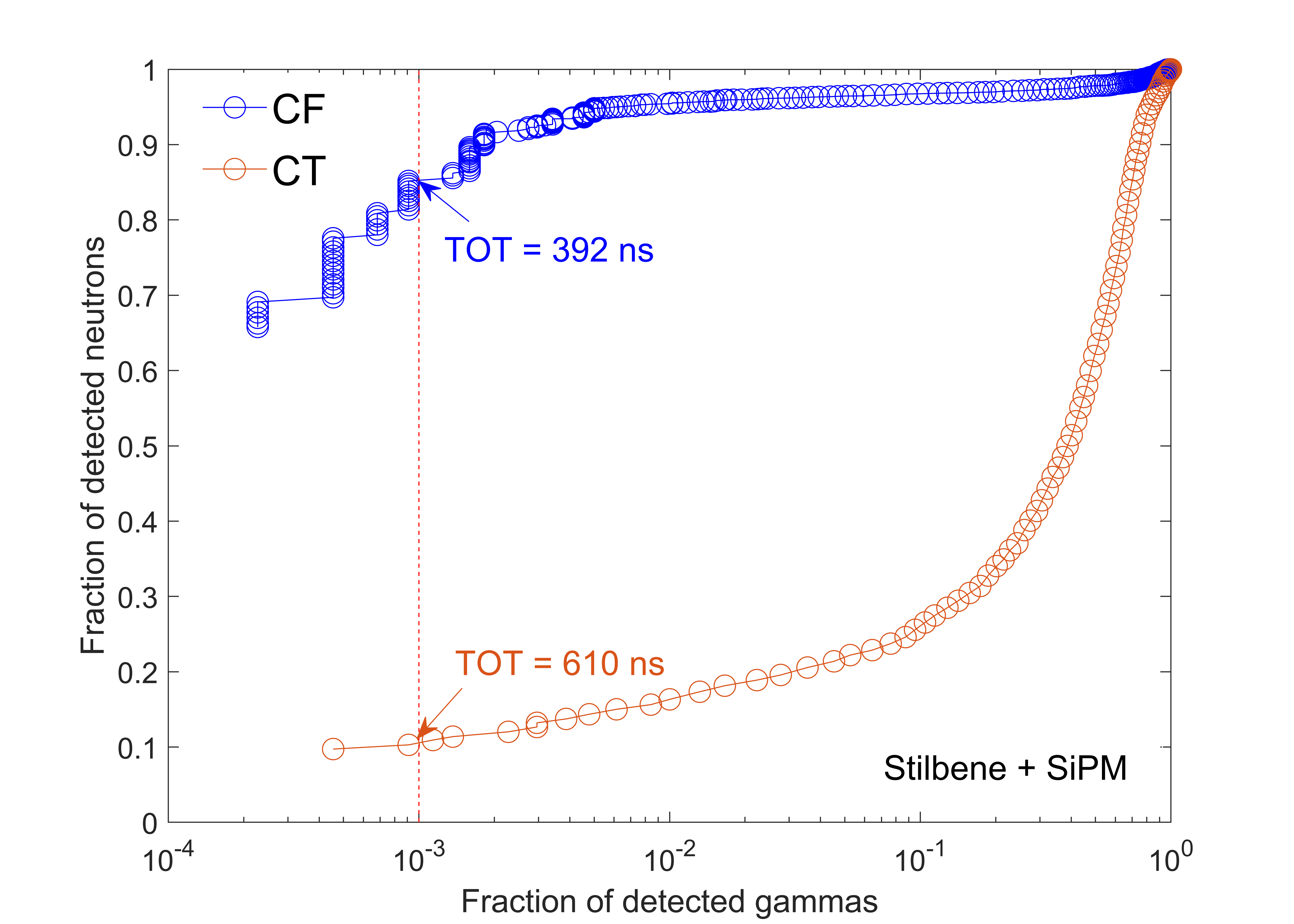}
\caption{\label{fig:xi} ROC curves for the three configurations. Top - LS+PMT, center - Stilbene+PMT, bottom - Stilbene+SiPM. The red line indicates the level of accepted fraction of gamma rays. The TOT threshold corresponding to this fraction is shown for the CF-TOT and the CT-TOT methods.}
\end{figure}

\noindent In order to compare the CT-TOT method to the CF-TOT method quantitatively, we used Receiver Operator Characteristic (ROC)~\cite{10.1093/clinchem/39.4.561} curves to study the gamma-ray rejection capability of the methods rather than the standard figure-of-merit (FOM). In the ROC curves approach, one plots the fraction of detected neutrons (true positive) as a function of the fraction of accepted gamma-rays (false positive) for a given TOT threshold value. A good separation would mean a high fraction of measured neutrons for nearly complete rejection of gamma-ray pulses. The ROC curves were generated by moving the TOT value threshold on the TOT vs total light output plots of Figure \ref{fig:vii} and Figure \ref{fig:viii} and determining the percent of gamma-rays and neutrons above this threshold. Figure \ref{fig:xi} shows the ROC curves for the three configurations. For the purpose of comparison, we select a desired gamma-ray acceptance value (vertical dashed red line) and determine the percent of detected neutrons. The values of TOT thresholds which correspond to this gamma-ray acceptance fraction is indicated for each of the TOT methods.
\noindent For CT-TOT, a high rejection of gamma-rays ($<$1$\times$10$^{-3}$) is possible at the cost of losing more than 80$\%$ of the neutrons. In comparison, CF-TOT provides a much more satisfactory PSD performance, namely, detection of $>$80$\%$ of neutrons for the same rejection rate of gamma-rays.

\section{Summary and discussion}

In this article, we performed a quantitative comparison between the Constant Fraction TOT neutron/gamma-ray PSD and the Constant Threshold TOT method. The comparison was performed for three scintillator-light detector configurations: liquid scintillator coupled to PMT, stilbene crystal coupled to PMT and stilbene crystal coupled to SiPM. For the comparison, we used digitized pulses obtained by irradiating the detectors with an Am-Be neutron source. The pulses were classified as neutrons and gamma-rays using the widely accepted and tested charge comparison method.
Due to the fact that in the Constant-Threshold-TOT frequency distribution there is a significant overlap between the gamma-ray and the neutron TOT and it does not exhibit the standard two well separated peaks, we used ROC curves to study quantitatively the gamma-ray rejection capability of the methods rather than the conventional figure-of-merit. 
Based on our studies we can conclude the following:

\begin{itemize}
\item For CT-TOT, a reasonable rejection of gamma-rays (accepted fraction $<$10$^{-3}$) occurs at the expense of rejecting 60-90$\%$ of neutrons.
\item The rejected events are all below 1000-1500 keVee, thus CT-TOT is not useful as PSD method for low energy neutrons.
\item Contrary to that, CF-TOT allows the same gamma rejection at the expense of losing only10-20$\%$ of neutrons.
\item CF-TOT method is useful for separating neutrons and gamma-rays events down to 100 keVee. 
\end{itemize}

\noindent Clearly, by comparing the frequency distributions of CF-TOT (Figure \ref{fig:x}) to those the CC (Figure \ref{fig:vi}) it appears that the CF-TOT is inferior to the CC method, since the wings are much broader. This is expected, because contrary to the charge comparison method, which is robust due to its integrative nature, the TOT method is more sensitive to signal noise and local irregularities. Therefore, especially for the PMT pulses, some degree of pulse smoothing or integration is advisable. Here we applied a ten points smoothing average. In both TOT cases, the noise magnitude controls the choice of the threshold voltage or the fraction. \\
Our aim is to avoid pulse digitization and to perform CF-TOT PSD using a simple analog circuitry suitable for multichannel ASIC implementation. As far as we are aware, CF-TOT electronics has not been implemented yet. The conventional CF discriminator provides a CF trigger on the leading edge of the pulse, but not on both sides of the pulse. The leading edge CFD has been implemented in the 128 channels VFAT3 front-end ASIC~\cite{Abbaneo_2016}. The CT-TOT has been implemented in the 64-channel TOFPET2 ASIC~\cite{Rolo_2013,Bugalho_2019}. 
We are presently aiming to develop a circuit based on a principle similar to the Constant Fraction Discriminator (CFD)~\cite{GEDCKE1967377}, where a pulse is passively attenuated by an apriori selected fraction and is compared to itself, thus an internal normalization is performed.

\acknowledgments
The authors would like to thank Dr. Shikma Bressler from the Department of Particle Physics $\&$ Astrophysics of the Weizmann Institute of Sciences for lending us the Stilbene scintillator and Prof. Amos Breskin for his useful comments in preparing the manuscript. The authors would also like to thank Anatoly Rodnianski from the Unit of Nuclear Engineering of Ben-Gurion University of the Negev for his assistance in handling the radiation sources and Doron Bar from Soreq Nuclear Research Center for helpful discussions related to the data analysis. The work was performed under grant no. 3-16313 from the Israel Ministry of Science and Technology.

\bibliographystyle{JHEP}
\bibliography{bibliography}

\end{document}